\pdfoutput=1 
\newif\ifextended\extendedtrue
\documentclass[sigconf, nonacm]{acmart}
\usepackage{graphicx}
\usepackage{subcaption}  
\usepackage{tikz}
\usetikzlibrary{patterns,matrix,arrows.meta, shapes.geometric, fit, positioning, calc, backgrounds, decorations.pathreplacing}
\usepackage{xcolor}
\usepackage[capitalise]{cleveref}

\usepackage{balance}
\usepackage{pgfplots}
\pgfplotsset{compat=newest}         
\usepackage{amsmath}       
\usepackage{amssymb}       
\usepackage{enumitem}      
\usepackage{verbatim}     
\usepackage{caption}       
\usepackage{algorithm}    
\usepackage{algorithmic}   

\begin{document}

\title{FluxSieve: Unifying Streaming and Analytical Data Planes for Scalable Cloud Observability}

\author{Adriano Vogel}
\orcid{0000-0003-3299-2641}
\affiliation{%
    \institution{Dynatrace Research}%
    \city{Linz}%
    \country{Austria}%
}
\email{adriano.vogel@dynatrace.com}

\author{Sören Henning}
\orcid{0000-0001-6912-2549}
\affiliation{%
  \institution{Dynatrace Research}%
  \city{Linz}%
  \country{Austria}%
}
\email{soeren.henning@dynatrace.com}

\author{Otmar Ertl}
\orcid{0000-0001-7322-6332}
\affiliation{%
    \institution{Dynatrace Research}%
    \city{Linz}%
    \country{Austria}%
}
\email{otmar.ertl@dynatrace.com}

\begin{abstract}
Despite many advances in query optimization, indexing techniques, and data storage, modern data platforms still face difficulties in delivering robust query performance under high concurrency and computationally intensive queries. This challenge is particularly pronounced in large-scale observability platforms handling high-volume, high-velocity data records. For instance, recurrent, expensive filtering queries at query time impose substantial computational and storage overheads in the analytical data plane. 

In this paper, we propose FluxSieve, a unified architecture that reconciles traditional pull-based query processing with push-based stream processing by embedding a lightweight in-stream precomputation and filtering layer directly into the data ingestion path. This avoids the complexity and operational burden of running queries in dedicated stream processing frameworks.
Concretely, this work (i) introduces a foundational architecture that unifies streaming and analytical data planes via in-stream filtering and records enrichment, (ii) designs a scalable multi-pattern matching mechanism that supports concurrent evaluation and on-the-fly updates of filtering rules with minimal per-record overhead, (iii) demonstrates how to integrate this ingestion-time processing with two open-source analytical systems---Apache Pinot as a Real-Time Online Analytical Processing (RTOLAP) engine and DuckDB as an embedded analytical database, and (iv) performs comprehensive experimental evaluation of our approach. 

Our evaluation across different systems, query types, and performance metrics shows up to orders-of-magnitude improvements in query performance at the cost of negligible additional storage and very low computational overhead.
These results indicate that shifting stable, high-cost filtering operations from the analytical plane into the streaming data plane is an effective design pattern for large-scale data analytics, enabling simple but efficient architectures. The proposed approach is especially relevant to practitioners working on industry-grade analytics platforms and to database and stream processing system designers.
\end{abstract}

\maketitle
\pagestyle{plain}
\section{Introduction}

Query performance optimization in large-scale data processing systems---including traditional databases, data warehouses, and modern data lakehouses---is a fundamental challenge in data management. Decades of work on query plan optimization, indexing strategies, partitioning schemes, and materialized views have significantly improved performance~\cite{armbrust2021}. However, achieving consistently low latency remains difficult for massive volumes of semi-structured and unstructured data such as system logs, application traces, and event streams. The difficulty stems from multiple factors: the scale of data that must be scanned or filtered, the complexity of analytical queries, trade-offs between storage efficiency and query speed, and the distributed nature of modern architectures.

Even with advances in columnar storage formats, indexing, adaptive query execution, and cost-based optimizers, log data remains especially challenging due to its high ingestion rates. Maintaining sub-second query response times over petabyte-scale log datasets continues to push system designs to their limits and motivates new approaches to efficient data processing~\cite{abadi2020}.

Conventional database and data processing architectures typically embody a self-centered, \emph{pull-based} query model. Users and applications repeatedly poll centralized repositories to retrieve the current state on demand, assuming that data is relatively static and that query latency is acceptable. While this approach has proven effective for many transactional and analytical workloads, it is ill-suited to modern high-velocity environments such as real-time analytics and event-driven architectures. Pull-based querying requires traversing indexes and executing query plans on each request, consuming computational resources and often re-reading data that has not changed since the last poll. It also tightly couples databases to clients by requiring the database to sustain all query loads, limiting scalability and increasing latency. As a result, pull-based systems suffer from repeated polling, increased resource consumption, and delayed insights. 

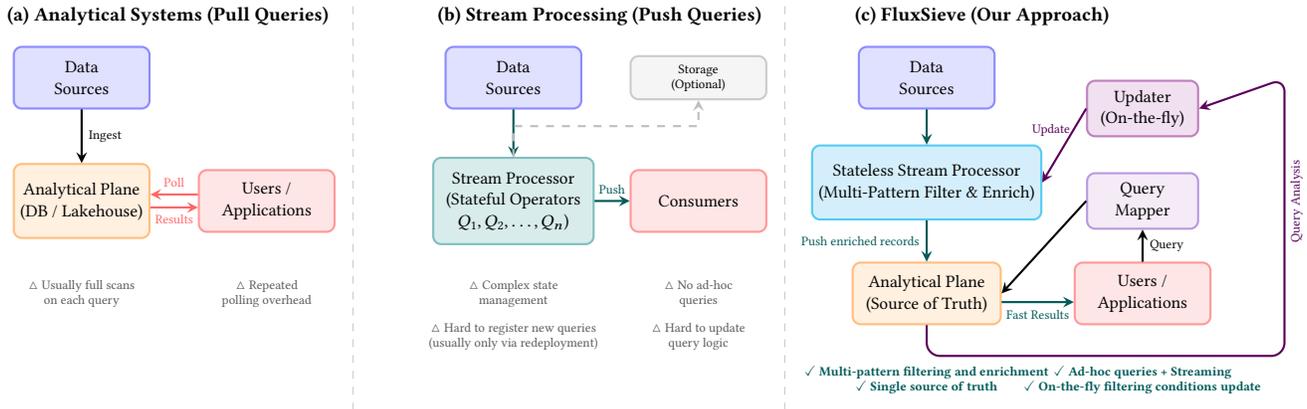
\begin{figure*}
\centering
\begin{tikzpicture}[
    scale=0.82,
    every node/.style={transform shape},
    node distance=0.8cm,
    block/.style={
        rectangle,
        draw,
        thick,
        rounded corners=3pt,
        minimum width=2.2cm,
        minimum height=1.0cm,
        align=center,
        font=\small
    },
    warning/.style={font=\scriptsize, text=black!65},
    benefit/.style={font=\scriptsize\bfseries, text=teal!70!black},
    arrow/.style={->, thick, >=stealth},
    poll/.style={->, thick, >=stealth, red!60},
    push/.style={->, thick, >=stealth, teal!70!black},
    feedback/.style={->, thick, >=stealth, violet!70!black}
]

\begin{scope}[xshift=0cm]
    \node[font=\bfseries] at (3.2, 5.0) {(a) Analytical Systems (Pull Queries)};
    
    \node[block, fill=blue!12, draw=blue!45] (src-a) at (1.8, 4.0) {Data\\Sources};
    
    \node[block, fill=orange!12, draw=orange!50, minimum height=1.2cm] (db-a) at (1.8, 2.0) {Analytical Plane\\(DB / Lakehouse)};
    
    \node[block, fill=red!10, draw=red!40] (user-a) at (4.8, 2.0) {Users /\\Applications};
    
    \draw[arrow] (src-a) -- node[right, font=\scriptsize] {Ingest} (db-a);
    
    \draw[poll] ([yshift=3pt]user-a.west) -- ([yshift=3pt]db-a.east)
        node[midway, above, font=\scriptsize, text=red!60] {Poll};
    
    \draw[poll] ([yshift=-3pt]db-a.east) -- ([yshift=-3pt]user-a.west)
        node[midway, below, font=\scriptsize, text=red!60] {Results};
    
    \node[warning, align=center] at (1.8, 0.5) {$\triangle$ Usually full scans\\on each query};
    \node[warning, align=center] at (4.8, 0.5) {$\triangle$ Repeated\\polling overhead};
\end{scope}

\begin{scope}[xshift=7.0cm]
    \node[font=\bfseries] at (3.2, 5.0) {(b) Stream Processing (Push Queries)};
    
    \node[block, fill=blue!12, draw=blue!45] (src-b) at (1.8, 4.0) {Data\\Sources};
    
    \node[block, fill=teal!15, draw=teal!50, minimum height=1.4cm, minimum width=2.6cm] (stream-b) at (1.8, 2.0) {Stream Processor\\(Stateful Operators\\$Q_1, Q_2, \ldots, Q_n$)};
    
    \node[block, fill=red!10, draw=red!40] (consumer-b) at (4.8, 2.0) {Consumers};
    
    \node[block, fill=gray!8, draw=gray!40, minimum height=0.7cm, font=\scriptsize] 
        (storage-b) at (4.8, 4.0) {Storage\\(Optional)};
    
    \draw[push] (src-b) -- (stream-b);
    \draw[push] (stream-b.east) -- (consumer-b.west) 
        node[midway, above, font=\scriptsize, text=teal!70!black] {Push};
    \draw[arrow, gray!50, dashed] (stream-b.north) -- ++(0, 0.5) -| (storage-b.south);
    
    \node[warning, align=center] at (1.8, 0.5) {$\triangle$ Complex state\\management};
    \node[warning, align=center] at (4.8, 0.5) {$\triangle$ No ad-hoc\\queries};
    \node[warning, align=center] at (1.8, -0.2) {$\triangle$ Hard to register new queries \\ (usually only via redeployment)};
    \node[warning, align=center] at (4.8, -0.2) {$\triangle$ Hard to update\\query logic};
\end{scope}

\begin{scope}[xshift=14.0cm]
    \node[font=\bfseries] at (2.4, 5.0) {(c) FluxSieve (Our Approach)};
    
    \node[block, fill=blue!12, draw=blue!45] (src-c) at (1.5, 4.0) {Data\\Sources};
    
    \node[block, fill=cyan!15, draw=cyan!55, thick, minimum width=2.4cm, minimum height=1.2cm] (flux-c) at (1.5, 2.3) {Stateless Stream Processor\\(Multi-Pattern Filter \& Enrich)};
    
    \node[block, fill=orange!12, draw=orange!50, minimum width=2.4cm, minimum height=1.0cm] (db-c) at (1.5, 0.5) {Analytical Plane\\(Source of Truth)};
    
    \node[block, fill=violet!12, draw=violet!50, minimum height=0.9cm, minimum width=1.8cm] 
        (updater-c) at (5.0, 3.5) {Updater\\(On-the-fly)};
    
    \node[block, fill=blue!8!violet!8, draw=blue!40!violet!40, minimum height=0.9cm, minimum width=1.8cm] 
        (mapper-c) at (5.0, 2.0) {Query\\Mapper};
    
    \node[block, fill=red!10, draw=red!40] (user-c) at (5.0, 0.5) {Users /\\Applications};
    
    \draw[push] (src-c) -- (flux-c);
    
    \draw[push] (flux-c) -- (db-c) 
        node[midway, left, font=\scriptsize] {Push enriched records};
    
    \draw[feedback] (updater-c.west) -- (flux-c.east)
        node[midway, above, font=\scriptsize, xshift=-6pt, yshift=1pt] {Update};
    
    \draw[feedback, rounded corners=5pt] (db-c.south) -- ++(0, -0.5) -- ++(5.8, 0) -- ++(0, 4.5) -- (updater-c.east);
    
    \node[font=\scriptsize, text=violet!70!black, rotate=90] at (7.5, 2.0) {Query Analysis};
    
    \draw[arrow] (user-c.north) -- (mapper-c.south)
        node[midway, right, font=\scriptsize] {Query};
    
    \draw[arrow] (mapper-c.west) -- (db-c.east);
    
    \draw[push] ([yshift=-4pt]db-c.east) -- ([yshift=-4pt]user-c.west)
        node[midway, below, font=\scriptsize, text=teal!70!black] {Fast Results};
    
    \node[benefit, align=center] at (1.5, -0.9) {$\checkmark$ Multi-pattern filtering and enrichment \\$\checkmark$ Single source of truth};
    \node[benefit, align=center] at (5.0, -0.9) {$\checkmark$ Ad-hoc queries + Streaming \\$\checkmark$ On-the-fly filtering conditions update};
\end{scope}

\draw[gray!40, dashed] (6.2, -1.4) -- (6.2, 5.4);
\draw[gray!40, dashed] (13.2, -1.4) -- (13.2, 5.4);

\end{tikzpicture}
\caption{Query processing paradigms: (a)~Traditional pull-based queries; (b)~Stream Processing; (c)~FluxSieve unifies by performing in-stream multi-pattern filtering and enrichment while preserving the analytical plane as the source of truth.}
\label{fig:approach-comparison}
\end{figure*}

Stream processing addresses these limitations by introducing push-based queries, which enable real-time responsiveness and eliminate repetitive polling. In this model, data flows continuously through processing pipelines, and results are \emph{pushed} to consumers immediately. However, pure stream processing introduces its own challenges. First, designing and operating streaming systems often requires complex state management and careful handling of fault tolerance and scalability. Second, the lack of readily accessible, persistent state in many streaming engines complicates ad-hoc analysis and historical data access, limiting their suitability for workloads that combine streaming and batch processing~\cite{Fragkoulis2023}. Third, traditional stream processing frameworks such as Apache Flink, Apache Spark Structured Streaming, and Kafka Streams typically demand substantial engineering effort to develop, deploy, and maintain production systems. Finally, running separate stream processors can increase both infrastructure costs and operational overhead~\cite{SEAA2023}.

To balance these trade-offs, recent research and industrial systems have explored \emph{hybrid architectures} that integrate streaming push mechanisms with traditional database pull queries~\cite{dulay2025}. These architectures aim to combine the immediacy of streaming with the flexibility of on-demand querying, supporting both continuous and interactive workloads~\cite{Grandi2022}. This hybrid model is particularly attractive in domains such as large-scale observability systems, where real-time insights must coexist with historical analysis.

This work advances this line of research by unifying streaming and database paradigms through an efficient in-stream precomputing and filtering mechanism that operates directly within the data ingestion pipeline. Our approach, FluxSieve, embeds lightweight, stateless stream-processing capabilities into the data path between sources and analytical processing architectures. Records can thus be filtered and enriched in-stream before they reach the analytical storage layer, persisting data in optimized mode to improve data retrieval, pruning, and downstream query performance. \Cref{fig:approach-comparison} overviews our approach compared to other paradigms.

A key element of our approach is efficient multi-pattern matching, which allows multiple filtering conditions to be evaluated against incoming records. By performing pattern matching and enrichment at ingestion time, we shift computational work upstream, removing expensive filtering operations from the analytical plane. This in-stream filtering and enrichment serves dual purposes: it reduces storage and index maintenance costs, and it precomputes derived filtering attributes that would otherwise require costly query-time computation. Consequently, queries in the analytical plane operate on a refined, enriched dataset optimized for analytical access patterns, yielding substantial gains in performance and overall system efficiency. Our unified architecture preserves the benefits of stream-oriented thinking while avoiding the complexity and overhead of traditional streaming frameworks, moving expensive and recurrent filtering queries out of the database.\footnote{This paper focuses on moving filtering to the streaming data plane, though suitable aggregations could likewise be precomputed in-stream to reduce analytical query load.}
This paper provides the following contributions:

\begin{itemize}

\item We propose FluxSieve as a foundational architecture to unify the analytical and streaming data planes.
\item We design an efficient in-stream filtering approach based on multi-pattern matching for stream processing systems.
\item We present an approach for on-the-fly updating of filtering conditions in stream processors.
\item We demonstrate how to integrate FluxSieve with two open-source analytical processing systems: Apache Pinot~\cite{pinot2018} as a Real-Time Online Analytical Processing (RTOLAP) system and DuckDB~\cite{raasveldt2019} as an embedded analytical database.
\item We provide comprehensive experimental evaluation of our approach across diverse systems, queries, and metrics.
\end{itemize}

This article is structured as follows. \Cref{sec:background} presents the background and motivation. \Cref{sec:approach} describes our approach in detail. \Cref{sec:evaluation} discusses the evaluation methodology, and \Cref{sec:evaluation-olap,sec:evaluation-db} present the experimental results. \Cref{sec:related} overviews the state of the art. Finally, \Cref{sec:closing} concludes this paper.

\section{Background and Motivation}\label{sec:background}
This section introduces our motivation use-case~\cref{sec:motivation} and \cref{sec:challenges} expands the description of existing challenges.

\subsection{Real-world Use-case and Motivation}\label{sec:motivation}
Modern cloud–native systems emit a heterogeneous stream of \emph{observability signals} \textbf{logs}, distributed \textbf{traces} (collections of \emph{spans}), high-resolution \textbf{metrics}, execution \textbf{profiles}, and \textbf{events}~\cite{ICPE2024}. These data feed two dominant classes of analytical queries:

\begin{enumerate}
\item \textit{Recurrent dashboards and alerts}. Operators continuously recompute service-level indicators such as request latency percentiles, error rate time series, and resource utilization. 

\item \textit{Periodic forensics and capacity analyses}. Post-incident investigations, threat-hunting, regression triage, or long-term capacity planning often demand full scans of weeks of raw telemetry, complex pattern matching over free-text messages, or multi-attribute joins between logs, traces, and metrics. These queries are computationally intensive and time-sensitive, yet they are also highly repetitive, e.g., executing similar filters periodically across incidents.
\end{enumerate}

This work concentrates on \textbf{log} workloads exhibiting \emph{very high selectivity}, i.e.\ cases in which the answer consists of far fewer records than the input data. These scenarios most clearly expose the cost of exhaustive scanning and thus motivate improvements. Although logs are our primary focus, the proposed in-stream filtering and enrichment mechanism is general: it can pre-compute expensive, repetitive filters for a wide range of observability data types, including logs, spans, and metrics records.

High-severity incidents often demand ``needle-in-a-haystack’’ queries where from many millions of log lines, very few records match. The extreme selectivity and urgency of these workloads expose the performance ceiling of both pull-query analytics and naive streaming filters. Executing such ``needle-in-a-haystack’’ queries within operational time frames (seconds to minutes) is challenging because the analytical engine must scan many gigabytes or terabytes of semi-structured text and evaluate thousands of expressions. Furthermore, numerous observability queries are executed periodically using fixed filters across varying time windows.

\subsection{\textit{Status Quo} and Challenges}\label{sec:challenges}

Exploratory log analysis in production settings still relies predominantly on \emph{pull queries}: analysts or automated dashboards issue queries that are executed retrospectively over a large, immutable log corpus. This does not consider the continuous and fast arrival of new records (e.g., logs) that are usually relevant for serving queries. Moreover, although state-of-the-art systems mitigate brute-force scans through inverted, range, and full-text indexes, several bottlenecks endure.

\subsubsection{Limitations of Pull-Query Analytics}

\begin{enumerate}[label=(\alph*)]
  \item \textbf{Scan Overhead.}  
        Even with modern full-text search (FTS) indexes, highly selective queries usually demand reading and decompressing every candidate data segment that \emph{might} contain a match. The resulting I/O and CPU cycles dominate query execution times, limiting overall queries performance and system scalability.

  \item \textbf{Combinatorial Predicate Explosion.}  
        Operational playbooks frequently employ hundreds or thousands of string matching, regular expressions, or fuzzy matches. Encoding such composite conditions in traditional index structures is either infeasible (index bloat) or forces post-filter evaluation, negating index performance benefits.

    \item \textbf{FTS Index Limitations.}  
      FTS indexes are the state-of-the-art for string search on logs; however, they are inherently limited for scnearios of highly selective filter queries over massive data volumes. They consume substantial CPU during both construction and query execution, demand significant storage overhead, and query performance degrades as data grows into terabytes.

\end{enumerate}

\subsubsection{Stream Processing Potentials—and Its Drawbacks}

A natural counter-strategy is to shift complex filtering \emph{upstream} into stream processors, evaluating records predicates at ingestion. While conceptually appealing, this approach introduces its own challenges:

\begin{itemize}
  \item \textbf{Operational Fragmentation.}  
        Maintaining a separate, stateful streaming cluster (e.g., Flink, Kafka Streams) increases deployment, monitoring, and failure-handling complexity relative to a single converged analytics stack~\cite{DEBS2024}.

  \item \textbf{Performance and Efficiency.}  
        Efficient pattern matching and filtering in stream processing remain a significant challenge~\cite{Purtzel2025}. As rule sets expand, CPU and memory demands increase non-linearly, requiring aggressive resource provisioning that drives up operational costs. In typical stream processing architectures, such as the one demonstrated in ShuffleBench~\cite{ICPE2024}, aggregations are performed in separate stateful stream operators (processors). Beyond causing operational fragmentation, this separation increases the overhead required to update aggregation logic for new queries, particularly when compared to traditional pull-based queries executed on the analytical plane. Furthermore, despite recent advances in fault tolerance~\cite{DEBS2024}, these mechanisms continue to impose performance penalties. 

  \item \textbf{Rule-Set Agility.}  
        Updating hundreds of filtering expressions typically requires code changes, recompilation, and redeployment, delaying the application of fresh data when rapid iteration is most critical.
\end{itemize}

To overcome the aforementioned limitations and challenges, we propose an \emph{unified approach} that fuses the strengths of streaming and analytical systems: complex pattern matching and semantic enrichment are performed \emph{in-flight}, yet the augmented records become \emph{immediately} queryable in the analytical data plane. This approach is described in \Cref{sec:approach}.

\section{The FluxSieve Approach} \label{sec:approach}
Large-scale data processing systems—ranging from classical relational databases to modern Data Lakehouses—often face performance bottlenecks when executing recurrent and intensive queries over massive datasets. To address this challenge, we propose FluxSieve, an approach that introduces an \textbf{in-stream prefiltering and enrichment mechanism} prior to data ingestion into the \textit{analytical plane}. This method aims to reduce query latency and optimize resource utilization by selectively processing and augmenting records before they enter the analytical storage layer.

\subsection{Core Idea}

We \emph{turn the database inside out} by relocating targeted, high-value
computation from the analytical plane to the ingestion pathway, while
keeping the analytical plane as the authoritative \emph{source of truth}.
In-stream processing performs stateless prefiltering and lightweight
enrichment, aligned with observed workloads, transforming raw signals
into records that are primed for efficient query execution. The
analytical plane then interprets these enrichments (e.g., flags, compact
structures, partial aggregates) to accelerate query execution without
sacrificing correctness or completeness and without causing duplicated
data storage. This unification intends to preserve a clear logic:
\begin{itemize}
  \item \textbf{Authority:} The analytical plane remains canonical for storage, schema, and results. Enrichments are hints or accelerators, not substitutes for correctness.
  \item \textbf{Performance:} Recurrent or costly operations (filtering) are precomputed in-stream, shrinking candidate sets and lowering query-time CPU and I/O.
  \item \textbf{Adaptation:} The \textbf{Query Profiler} (analytical plane) detects expensive and frequent queries, and registers conditions via the \textbf{Updater Component} for in-stream compilation and multi-pattern matching (e.g., Hyperscan~\cite{wang2019}).
\end{itemize}

The approach operates between the \textit{data sources} and the \textit{analytical plane}, leveraging a stream processor to perform key operations:
\begin{itemize}
    \item \textbf{Prefiltering:} Matches records that are relevant to queries.
    \item \textbf{Enrichment:} Augment relevant records with lightweight, storage-efficient metadata that encodes query relevance. These enrichment fields are designed to minimize storage overhead while remaining highly compressible under columnar encoding schemes (e.g., run-length encoding), making them particularly effective for column-oriented storage systems (e.g., analytical databases, lakehouses).
\end{itemize}

\subsection{Architecture}
The architecture consists of the following modules:
\begin{enumerate}
    \item \textbf{Data Source Interface:} Captures raw records from heterogeneous sources (e.g., event streams, log aggregators).
    \item \textbf{Stream Processor:} Applies filtering and enrichment logic in real time using efficient multi-pattern matching, annotating records with precomputed query results.
    \item \textbf{Updater Component:} Synchronizes filtering conditions and enrichment rules with evolving query patterns, triggering pattern matching engine recompilation as needed.
    \item \textbf{Analytical Plane:} Storage and query execution layer, leveraging enriched metadata for efficient query execution. It \emph{embeds the \textbf{Query Profiler}} to analyze queries and execution plans, whereas expensive/frequent filtering conditions are moved for in-stream processing.
    \item \textbf{Query Mapper:} Translates incoming queries into optimized internal queries that exploit the precomputed fields. The query mapper enables the analytical plane to bypass expensive full-table scans and predicate evaluations.
\end{enumerate}

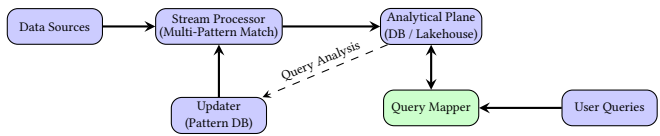
\begin{figure}[h!]
\centering
\begin{tikzpicture}[
    node distance=1.0cm and 1.4cm, 
    auto,
    >=stealth,
    scale=0.5,
    every node/.style={transform shape},
    block/.style={
        rectangle,
        draw,
        fill=blue!20,
        rounded corners,
        minimum height=3em,
        minimum width=8.0em,
        align=center,
        font=\large   
    },
    mapper/.style={
        rectangle,
        draw,
        fill=green!20,
        rounded corners,
        minimum height=3em,
        minimum width=8.0em,
        align=center,
        font=\large   
    }
]

\node[block] (source) {Data Sources};
\node[block, right=of source] (stream) {\shortstack{Stream Processor \\ (Multi-Pattern Match)}};
\node[block, below=1.4cm of stream] (updater) {\shortstack{Updater \\ (Pattern DB)}};
\node[block, right=2.6cm of stream] (dps) {\shortstack{Analytical Plane \\ (DB / Lakehouse)}};
\node[mapper, below=1.2cm of dps] (mapper) {Query Mapper};
\node[block, right=2.2cm of mapper] (user) {User Queries};

\draw[->, thick] (source) -- (stream);
\draw[->, thick] (stream) -- (dps);
\draw[->, thick] (updater) -- (stream);
\draw[->, dashed] (dps) -- (updater)
    node[midway, above, sloped] {\large Query Analysis};
\draw[->, thick] (user) -- (mapper);
\draw[<->, thick] (mapper) -- (dps);

\end{tikzpicture}
\caption{Architecture of our approach.}
\label{fig:tikz-architecture}
\end{figure}

\subsection{In-Stream Multi-Pattern Matching}\label{sec:matching}

A critical requirement for the stream processor is the ability to evaluate \textit{many filtering conditions simultaneously} against each incoming record with minimal latency overhead. Naively iterating through hundreds or thousands of filter predicates introduces prohibitive per-record costs, undermining the benefits of in-stream processing.

Therefore, the stream processor leverages \textbf{multi-pattern matching algorithms} to achieve efficient matching at scale. We implemented multi-pattern matching using \textbf{Hyperscan}~\cite{wang2019},\footnote{Java bindings available at \url{https://github.com/gliwka/hyperscan-java}} which is a high-performance regular expression matching library (originally developed by Intel) that compiles multiple patterns into an optimized finite automaton. Hyperscan evaluates all registered patterns in a single pass over the input data, supporting matching of thousands of patterns with predictable, bounded latency per record, making it well-suited for stream-processing workloads.

The choice of matching engine can be tailored to the workload characteristics: regex-heavy workloads benefit from Hyperscan's compiled automata, while keyword-centric filtering may favor the Aho-Corasick algorithm~\cite{aho1975} or hash-based approaches. In this work, we opted for using Hyperscan as multi-string matching engine because it is a modern approach with appropriate algorithmic efficiency and support for productive implementations on cloud-native stream processors. Unlike approaches that use Hyperscan’s multi-pattern engine with a single pattern~\cite{giouroukis2025}, we fully exploit it registering multiple patterns to efficiently filter streaming records.

\subsection{Dynamic Adaptation for On-the-Fly Filter Condition Updates}\label{sec:adaptation}

Despite significant advances in self-adaptation for parallelism and elasticity in stream processing executions~\cite{vogel2022}, there remains a lack of approaches that support on-the-fly self-adaptation of stream processors’ logic without requiring expensive application redeployments. In this work, we are interested in updating on-the-fly stream processors that continuously execute filtering conditions over incoming records. This section describes our approach and its integration into a distributed stream-processing framework.

A monitoring module within the analytical plane can identify \textit{queries of interest} based on workload analysis—detecting frequently executed queries, recurring filter patterns, and high-cost query segments. Filtering conditions and enrichment schemas are propagated to the stream processor via the updater component. This feedback loop ensures that the stream processing logic remains aligned with query requirements.

The updater component maintains a \textbf{pattern matching engine} that encapsulates all active filtering conditions and their associated enrichment rules. When new queries of interest are identified or existing patterns become obsolete, the system performs the following steps to update the matching engine:

\begin{enumerate}
    \item \textbf{Delta Computation:} The updater computes the difference between the current pattern set and the target pattern set, identifying additions, modifications, and removals.
    \item \textbf{Pattern Matching Engine Recompilation:} For engines like Hyperscan, the pattern matching engine is recompiled into an optimized automaton. This compilation step—while computationally intensive—is performed asynchronously and does not block ongoing stream processing.
    \item \textbf{On-the-fly Rules Update:} Once the new pattern matching engine is ready, it is swapped into the active stream processor. In-flight records continue to be processed against the previous matching engine until the swap completes, ensuring no records are incorrectly filtered during transitions.
    \item \textbf{Consistency Propagation:} Newly ingested records with updated enrichment fields are correctly interpreted at query time as the query mapper is notified of the updated schemas.
\end{enumerate}

This architecture enables \textbf{continuous evolution} of the filtering logic without service interruption, allowing the system to adapt to shifting query patterns, onboard new analytical use cases, and deprecate stale filters while avoiding stream processing downtime.

\subsubsection{Architecture Overview}

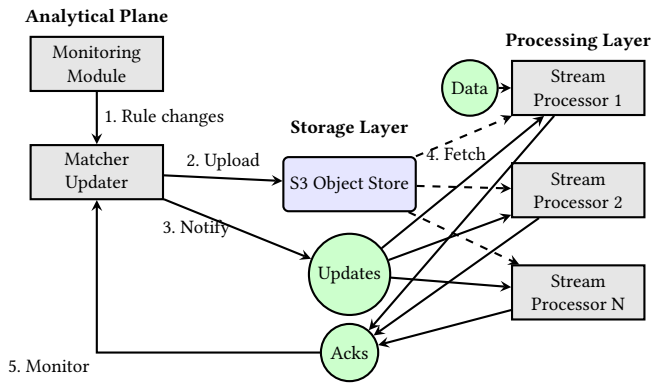
\begin{figure}[h!]
\centering
\begin{tikzpicture}[
    scale=0.8,
    every node/.style={transform shape},
    node distance=1.2cm,
    component/.style={
        rectangle,
        draw,
        thick,
        minimum width=2.2cm,
        minimum height=0.9cm,
        align=center,
        fill=gray!20,
        font=\normalsize
    },
    storage/.style={
        rectangle,
        rounded corners=2pt,
        draw,
        thick,
        minimum width=2.2cm,
        minimum height=0.9cm,
        align=center,
        fill=blue!10,
        font=\normalsize
    },
    kafka/.style={
        circle,
        draw,
        thick,
        minimum size=0.8cm,
        fill=green!20,
        font=\normalsize
    },
    arrow/.style={->, thick, >=stealth},
    dashed arrow/.style={->, thick, dashed, >=stealth},
    >=stealth
]

\node[component] (monitor) at (0,3.0) {Monitoring\\Module};
\node[component] (updater) at (0,1.2) {Matcher\\Updater};

\node[storage] (s3) at (4.2,1.0) {S3 Object Store};
\node[font=\normalsize\bfseries] at ([yshift=0.45cm]s3.north) {Storage Layer};

\node[component] (stream1) at (8.0,2.6) {Stream\\Processor 1};
\node[component] (stream2) at (8.0,0.9) {Stream\\Processor 2};
\node[component] (streamn) at (8.0,-0.8) {Stream\\Processor N};

\node[kafka] (updates) at (4.2,-0.5) {Updates};
\node[kafka] (acks)    at (4.2,-1.8) {Acks};
\node[kafka] (data)    at (6.2,2.6) {Data};

\draw[arrow] (monitor) -- node[right, font=\normalsize] {1. Rule changes} (updater);
\draw[arrow] (updater) -- node[above, font=\normalsize] {2. Upload} (s3);
\draw[arrow] (updater) -- node[left, font=\normalsize] {3. Notify} (updates);

\draw[arrow] (updates) -- (stream1);
\draw[arrow] (updates) -- (stream2);
\draw[arrow] (updates) -- (streamn);

\draw[dashed arrow] (s3) -- node[pos=0.42, below, font=\normalsize] {4. Fetch} (stream1);
\draw[dashed arrow] (s3) -- (stream2);
\draw[dashed arrow] (s3) -- (streamn);

\draw[arrow] (stream1) -- (acks);
\draw[arrow] (stream2) -- (acks);
\draw[arrow] (streamn) -- (acks);

\draw[arrow] (acks) -| node[below left, font=\normalsize] {5. Monitor} (updater);

\draw[arrow] (data) -- (stream1);

\node[above=0.05cm of monitor, font=\normalsize\bfseries] {Analytical Plane};
\node[above=0.05cm of stream1, font=\normalsize\bfseries] {Processing Layer};

\end{tikzpicture}
\caption{Pattern matching engine update: (1) detects changes, (2) updater compiles and uploads (3) notifications, (4) processors fetch the matching engine, (5) acknowledgments.}
\label{fig:update-flow}
\end{figure}

The pattern update mechanism follows a distributed coordination architecture designed to minimize latency, ensure consistency, and avoid disrupting active stream processing. Figure~\ref{fig:update-flow} illustrates the complete update lifecycle.

The architecture comprises four primary components:

\begin{itemize}
    \item \textbf{Filter Rules Management Interface:} Receives pattern updates from the analytical plane's monitoring module. This interface is expected to operate autonomously based on workload telemetry, query pattern analysis, and self-learning to detect frequent and expensive queries.
    
    \item \textbf{Matcher Updater:} The central orchestrator responsible for pattern matching engine compilation, versioning, and distribution. Upon receiving updated conditions, it compiles the patterns into optimized matching automata (e.g., a Hyperscan matching engine), assigns version tags, and coordinates the rollout across all stream processing instances.
    
    \item \textbf{Object Storage Layer (S3):}\footnote{It is important to note that Amazon S3 is used as an example of object storage technology supported in our approach. However, our approach could also be implemented in any other object stores with similar functional features.} Compiled pattern-matching engines are serialized and stored in a distributed object store. This design decision addresses two critical concerns: (1) large compiled matching engines (potentially tens to hundreds of megabytes) would create excessive overhead on Kafka topics, and (2) object storage provides cost-efficient, highly available distribution with built-in versioning.
    
    \item \textbf{Streaming Application (Matcher):} Each distributed stream processor instance maintains a local reference to the active pattern matching engine. Upon receiving update notifications via Kafka, instances asynchronously fetch the new compiled matching engine from object storage, confirm that the downloaded file matches the expected version, and validate its integrity using a stored checksum (e.g., a hash associated with the object) before performing a hot swap.
\end{itemize}

This architecture decouples resource-intensive \textit{compilation} from \textit{distribution} (decentralized, bandwidth-efficient), enabling scalable deployments and matching engine updates with low overhead.

\subsubsection{Pattern Matching Engine Update}
The update flow is:

\begin{enumerate}
    \item The monitoring module detects changes or is triggered by the Query Analyzer from Lakehouses or DBs and generates an updated set of filtering conditions.
    \item The Matcher Updater compiles the iltering conditions into a versioned pattern matching engine and uploads it to S3 with a unique version identifier.
    \item A notification message containing the version tag and S3 object reference is published to a dedicated Kafka topic.
    \item All streaming application instances consume the update notification, fetch the compiled pattern matching engine from S3, and prepare for activation.
    \item Upon retrieval and validation, each instance swaps the new pattern-matching engine into its active processing pipeline.
    \item Acknowledgment messages are sent back through Kafka to confirm successful matching engine activations across the cluster. This is an optional step.
\end{enumerate}

\subsubsection{Implementation on Kafka Streams}

The pattern update mechanism is implemented as a Kafka Streams application with a dual-topology design: one topology processes the primary data stream with active pattern matching, while the secondary topology handles pattern-matching engine updates via a dedicated control plane.

\paragraph{Communication Protocol}

The update coordination protocol leverages Kafka's exactly-once semantics and consumer group coordination to ensure consistent pattern activation across distributed instances. The Matcher Updater can subscribe to the acknowledgment topic to monitor rollout progress and detect any instances that fail to activate the new matching engine within a configurable timeout window.

\paragraph{Object Storage Distribution}

Rather than embedding compiled pattern-matching engines directly in Kafka messages, the system employs a \textbf{reference-based distribution model}. This approach provides several advantages:

\begin{enumerate}
    \item \textbf{Bandwidth Efficiency:} Kafka messages remain lightweight (typically under 1 KB), avoiding the throughput degradation associated with large message payloads. A compiled Hyperscan pattern matching engine for thousands of patterns can exceed 100 MB, which would overwhelm Kafka brokers if transmitted inline.
    
    \item \textbf{Cost Optimization:} S3 storage costs are significantly lower than Kafka retention costs for large binary objects. Additionally, S3 storage and lifecycle policies enable automatic archival of older pattern versions.
    
    \item \textbf{Caching:} Stream processors can leverage HTTP caching headers or similar services to reduce retrieval latency for geographically distributed deployments.
    
    \item \textbf{Versioning:} S3 object versioning ensures that each pattern matching engine version remains immutable and accessible, enabling rollback capabilities and audit trails.
\end{enumerate}

\paragraph{Kafka Streams Integration}

The update handling logic is implemented as a separate processor within the Kafka Streams topology, operating independently of the main data processing pipeline. This processor subscribes to the \texttt{matcher-updates} topic and triggers the update workflow without blocking record processing.
The integration leverages Kafka Streams' state stores to maintain metadata about active and pending pattern versions tracking:

\begin{itemize}
    \item Current active pattern version
    \item Pending update version (if an update is in progress)
    \item Activation timestamps for audit and monitoring
\end{itemize}

The data processing topology accesses the active pattern-matching engine via a shared, thread-safe reference. When there is a new pattern matching engine, the reference is updated, ensuring that subsequent records are processed with the latest patterns.

\section{Evaluation Method}\label{sec:evaluation}

This section describes the method to evaluate our proposed approach. In terms of the evaluation method, we adopted, when applicable and appropriate, the good practices, such as those from~\cite{raasveldt2018,hoefler2015}. We evaluated our approach (\cref{sec:approach}) against the baseline system. We consider as baseline the most appropriate approach for processing the tested queries; for instance, in Apache Pinot, the baseline is a full-text search (FTS) index because queries search string fields for terms. The comparison covers:

\begin{itemize}

\item Baseline -- Events ingested (Pinot) or written to Parquet;

\item FluxSieve (our approach) with in-stream filtering and enrichment -- Same workload, but filtering and enrichment of records with stream processor that flow to Pinot or Parquet.

\end{itemize}

\subsection {Metrics}

The following are the main metrics considered in our evaluation:

\begin{itemize}

\item Query performance: representative analytical queries over enriched vs. non-enriched datasets in different systems.

\item Ingestion throughput.

\item CPU resource usage.

\item Storage size: segment/file sizes.

\end{itemize}

The query performance evaluation uses the median over multiple runs as the main metric, and reports 95\% confidence intervals using bootstrapping by repeatedly resampling the measurements and recomputing the median as good practices~\cite{raasveldt2018,hoefler2015}.

The base queries used as workloads for evaluation are:
\begin{itemize}
\item Query 1: Filter with search on a string field for a non-matching term. 

\item Query 2: Filter with search on a string field for very rare matching condition. 

\item Query 3: Filter for a term condition on a string field and counting the number of matches (aggregation).

\item Query 4: Multi-field search. Search for records where two string fields contain arbitrary terms in their content.

\end{itemize}

\subsection{Queries Execution}\label{sec:execution}

In our experimental methodology, we explicitly distinguish between
\emph{cold} and \emph{hot} query executions. A \textbf{cold run} is
defined as an execution in which the underlying system is not expected
to benefit from data residing in RAM (Random Access Memory), cached
data, compiled query plans, or other warm-up effects.

To approximate this behavior, we apply cache-cleaning procedures before
cold runs, including: (i) clearing the operating system page cache
(\cref{sec:evaluation-db}); and (ii) redeploying the Kubernetes-based
infrastructure in which the analytical system under test is deployed
(\cref{sec:evaluation-olap}). These steps aim to remove data and
metadata retained from previous executions, ensuring that cold runs
simulate performance in a big data scenario where the required data and
query plans must be loaded from persistent storage because they are not
expected to fit in RAM.

In contrast, a \textbf{hot run} is an execution where the system caching mechanisms provided by the database engine, the operating system, and the surrounding infrastructure are maintained.
For hot runs, we do not perform cache-clearing steps between executions,
so that components such as buffer pools, OS page caches, and plan caches
can remain populated with data and compiled query plans from prior runs.
Because the timing characteristics of cold and hot runs can differ
substantially, we collect and report performance metrics for these two
modes separately. This separation allows us to capture both the
initial-cost behavior and when the data does not fit on RAM (cold runs) and the steady-state performance
(hot runs) of the evaluated system.

\subsection{Data Ingested and Queried}\label{sec:data}

In our experiments, we focus on query workloads over datasets ranging
from 5 million to 40 million records. The exact size
varies across setups and environments (e.g., different cluster
configurations or cloud deployments), but it is always within this
range. We consider this volume representative of realistic operational
datasets while also maintaining a reasonable balance between
experimental representativeness and cloud cost efficiency. For larger
data volumes, we expect the observed performance trends to hold, as the
underlying system behavior and bottlenecks are not fundamentally
altered by scaling beyond this range.

Each record has a logical schema representing observability logs: a \texttt{timestamp} field, which is a numerical value corresponding to the event time, a \texttt{status} field, an \texttt{eventType} field, and between 2 and 5 additional string \texttt{content} fields of size 60 words each field.
The number of \texttt{content} fields depends on the scenario.
We use at least two of fields to support multi-field queries (Query~4),
ensuring that our evaluation covers realistic search and filtering
patterns across multiple attributes. Additional \texttt{content} fields
are introduced in the scenario from~\cref{sec:evaluation-olap} for effective query runs on cold data, each query runs over a different field to avoid cached data and reduce the number of redeployments needed. This design allows us to stress and evaluate the system’s behavior under both cold and hot queries execution conditions over realistic and scalable deployments.

\section{Evaluation of Streaming Data Lake}\label{sec:evaluation-db}

In this section, we show the evaluation of our approach integrated into DuckDB~\cite{raasveldt2019}, which is an embedded analytical database optimized for high performance analytical workloads. Considering DuckDB's efficiency, evaluating our approach against DuckDB shows whether in-stream enrichment benefits embedded analytics scenarios. This scenario simulates a streaming data lake because ingested records are queryable with low latency as they are processed and written to files in near real time by a stream processor.
This experimental setup enables us to measure key aspects of our approach, including:
(I) scan reduction achieved through filter push-down on newly added attributes;
(II) query performance improvements compared to non-enriched files; and
(III) the potential overhead introduced by our approach relative to the baseline.

\subsection{Implementation and Setup}

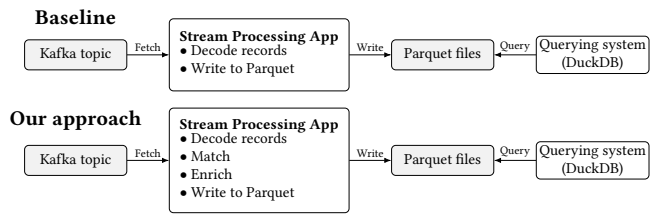
\begin{figure}[!h]
\centering
\resizebox{.48\textwidth}{!}{%
\begin{tikzpicture}[
  font=\huge,
  node distance=10mm and 14mm,
  box/.style={draw, rounded corners, align=center, minimum width=34mm, minimum height=10mm},
  stage/.style={draw, rounded corners, align=left, minimum width=56mm, minimum height=24mm, inner sep=3.5mm},
  data/.style={draw, rounded corners, align=center, minimum width=34mm, minimum height=10mm, fill=gray!10},
  arrow/.style={-Latex, thick}
]

\node[data] (kafka1) {Kafka topic};
\node[stage, right=of kafka1] (stream1) {\textbf{Stream Processing App}\\[-1mm]\begin{tabular}{@{}l@{}}
$\bullet$ Decode records\\
$\bullet$ Write to Parquet
\end{tabular}};
\node[data, right=of stream1] (parquet1) {Parquet files};
\node[box, right=of parquet1] (query1) {Querying system \\ (DuckDB)};

\draw[arrow] (kafka1) -- node[above]{\Large Fetch} (stream1);
\draw[arrow] (stream1) -- node[above]{\Large Write} (parquet1);
\draw[arrow] (query1) -- node[above]{\Large Query} (parquet1);

\node[above=4mm of kafka1, font=\fontsize{20}{24}\selectfont\bfseries] (baseline_label) {Baseline};

\node[data, below=25mm of kafka1] (kafka2) {Kafka topic};
\node[stage, right=of kafka2] (stream2) {\textbf{Stream Processing App}\\[-1mm]\begin{tabular}{@{}l@{}}
$\bullet$ Decode records\\
$\bullet$ Match\\
$\bullet$ Enrich\\
$\bullet$ Write to Parquet
\end{tabular}};
\node[data, right=of stream2] (parquet2) {Parquet files};
\node[box, right=of parquet2] (query2) {Querying system \\ (DuckDB)};

\draw[arrow] (kafka2) -- node[above]{\Large Fetch} (stream2);
\draw[arrow] (stream2) -- node[above]{\Large Write} (parquet2);
\draw[arrow] (query2) -- node[above]{\Large Query} (parquet2);

\node[above=4mm of kafka2, font=\fontsize{20}{24}\selectfont\bfseries] (our_label) {Our approach};

\end{tikzpicture}%
}
\caption{Comparison of deployments.}
\label{fig:arch-duckdb}
\end{figure}

Figure~\ref{fig:arch-duckdb} depicts two end-to-end pipelines that share the dataflow: \emph{Kafka $\rightarrow$ Stream Processor $\rightarrow$ Parquet $\rightarrow$ DuckDB}.
\textbf{Baseline (upper lane)} is a stream processing application consumes 10\,000 log events per second from a Kafka topic, deserializes the payloads and writes them as Parquet files. 
The resulting files expose only the native log attributes (\texttt{timestamp}, \texttt{status}, \texttt{content1}, \texttt{content2}, etc.). DuckDB later queries these Parquet files, benefiting from vectorization, projection and predicate push-down. We selected the optimized full scan as the baseline for DuckDB because it does not natively support full-text search (FTS) indexes. The current FTS index extension\footnote{\url{https://duckdb.org/docs/stable/core_extensions/full_text_search}} exhibited lower performance than the optimized full scan in our experimental scenario.

\textbf{Our approach (lower lane)} is an enhanced processor, deployed on the \emph{same} instance and fed at the \emph{same} records and rate, executes a suite of 1\,000 pattern-matching conditions over the fields.\footnote{There are two text fields filtered by a matching engine running in a stream processor.}
The stream processor emits an additional column \verb|matched_rule_ids INT[]| containing a \emph{sparse array} of identified matched. As DuckDB treats missing columns as \verb|NULL| and it supports efficient nested types, the array representation is, to the be best of our knowledge, the most appropriate and storage-efficient way to expose match metadata for local analytical workloads.

Keeping Kafka, Parquet formatting, compression settings, and the DuckDB query engine unchanged isolates the influence of enrichment strategy. Comparing the baseline against FluxSieve with filtering and the array-based enrichment allows us to quantify the effect of pre-computing match information on Parquet file size and query performance gains.

Both stream processors run on identical virtual machines provisioned with a single vCPU and 4 GiB RAM, maintaining the 10\,000 events/s ingest rate while producing 10 million events (10 minutes of traffic). 
For query-time benchmarks we copy the generated Parquet files and evaluate each workload under two hardware profiles:  
(i) the same single-core configuration used for ingestion, and  
(ii) a \texttt{4 vCPU / 16 GiB RAM} configuration, reflecting realistic analyst-facing scenarios where ad-hoc queries must scan large log volumes quickly. 
This dual-profile design aim at measuring how enrichment interacts with DuckDB’s intra-query parallelism and whether the benefits of pre-materialized \verb|matched_rule_ids| persist—or even amplify—when additional CPU resources are available.

\subsection{Overhead Analysis}

\begin{figure}[h!]
  \centering
  \includegraphics[width=\linewidth]{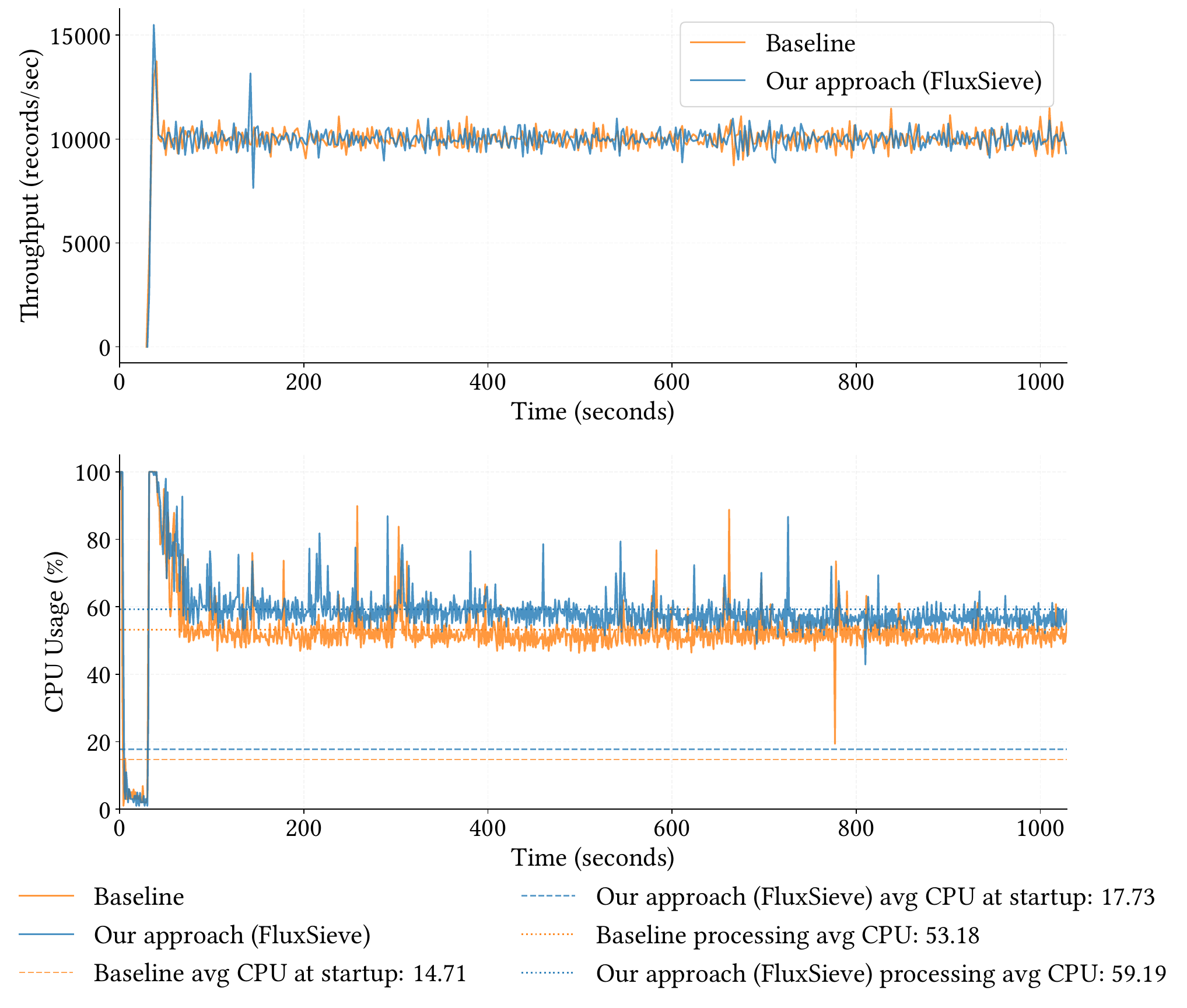}
  \caption{Overhead Analysis.}
  \label{fig:overhead}
\end{figure}

Figure~\ref{fig:overhead} compares the baseline DuckDB ingestion pipeline, which only decodes records and writes them to Parquet, with our extended pipeline that additionally performs matching and enrichment before writing to Parquet (Figure~\ref{fig:arch-duckdb}). Both setups were executed on the same single-core instance, with identical input rates of 10\,000 records per second and a 30-second warmup phase. The throughput curves for the two approaches are almost indistinguishable, indicating that the extra matching and enrichment logic in our approach does not introduce any noticeable lag or backpressure on the Kafka partitions and preserves comparable ingestion throughput.

The CPU usage measures the overhead of our approach's additional processing. Before data input starts, our approach uses, on average, 17.73\% CPU versus 14.71\% for the baseline, i.e., 20.5\% higher CPU usage than the baseline. This overhead is expected because our approach compiles the matching engines at startup. During steady-state processing, the baseline averages 53.18\% CPU versus 59.19\% for our approach, corresponding to 11.3\% lower CPU consumption for the baseline. This difference primarily reflects the additional work of filtering and enriching records in-stream.\footnote{The CPU frequency was fixed at 2.5 GHz for the experiments to reduce variability.}

Overall, these results show that our approach can sustain the ingestion throughput as the simpler baseline while incurring only a moderate increase in CPU usage.\footnote{We also measured the storage size of the Parquet files compressed with Zstandard (zstd), and no size differences were observed beyond double-digit precision.} Given that the extra CPU is traded for pre-filtered and enriched data that simplifies and accelerates downstream querying, this overhead is can be acceptable for deployments where query performance and efficiency are relevant.

\subsection{Query Performance}

The experiments used a dataset of 10 million log records as a representative workload. Each record contains the structured fields described in \cref{sec:data} and two text fields. This setup is intended to mimic realistic observability or logging data, where a mix of structured attributes and free-text fields is queried repeatedly. We consider two dimensions that are critical for analytical engines such as DuckDB:
(1) how the 10 million records are physically distributed across Parquet files (many small files vs. fewer larger files),
and (2) how much intra-query parallelism is available (single-core vs.\ four-core execution).  
For each combination, we evaluate two common query patterns:
(i) queries that \emph{return} all matching records (copy scenario), and
(ii) queries that \emph{count} matching records (aggregation scenario).  
These scenarios matter because file layout and parallelism are central configuration levers in data lake deployments, and returning vs.\ counting results represent two dominant usage modes (exploratory retrieval vs.\ analytical summarization).

\begin{figure}
    \centering
    \begin{subfigure}[b]{\linewidth}
        \centering
        \includegraphics[width=\linewidth]{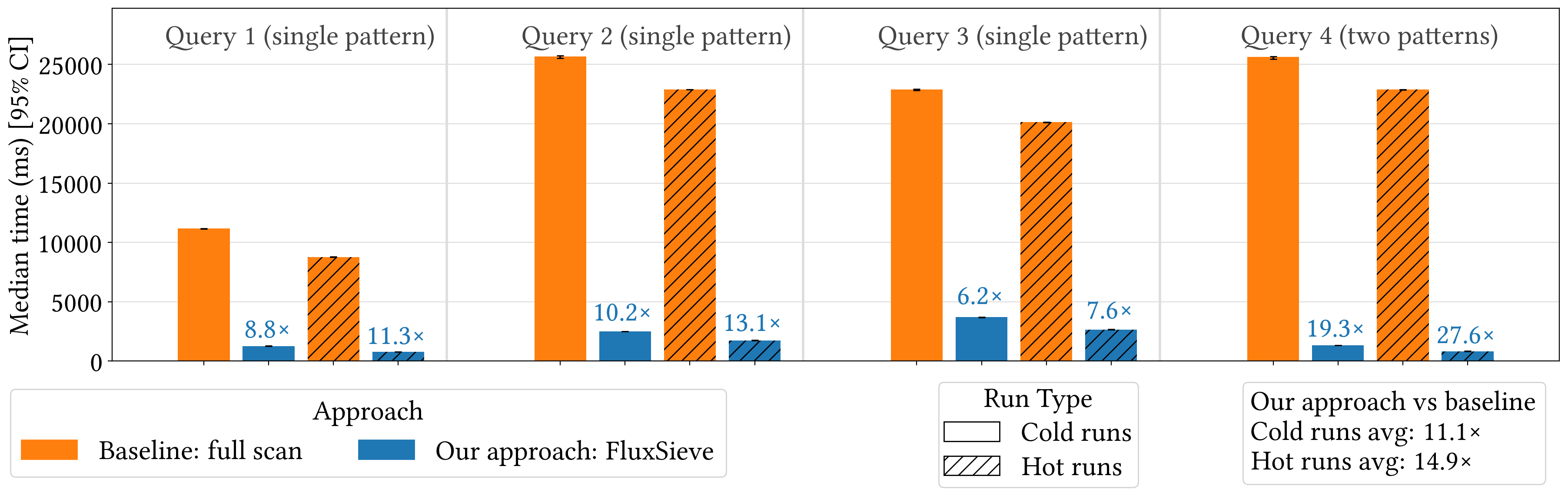}
        \caption{Returning records}
        \label{fig:copy_p1_5k-files}
    \end{subfigure}
    \vspace{0.5em}
    \begin{subfigure}[b]{\linewidth}
        \centering
        \includegraphics[width=\linewidth]{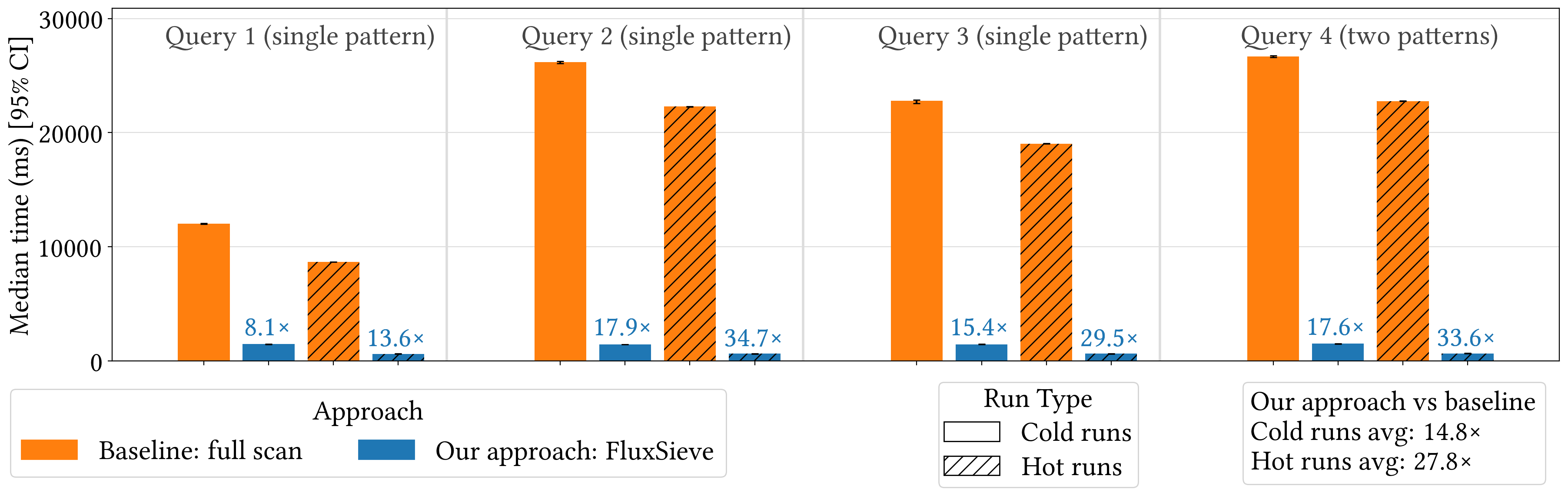}
        \caption{Counting records}
        \label{fig:count_p1_5k-files}
    \end{subfigure}
    \caption{\boldmath Parallelism level 1: $\approx$5k files, $\approx$2k records per file.}
    \label{fig:p1_5k-files}
\end{figure}

\paragraph{1 CPU core, $\approx$5k files, $\approx$2k records/file (\cref{fig:p1_5k-files}).}
With a single core and many small Parquet files, DuckDB needs to open and scan a large number of files to cover the 10 million records. In the \emph{copy} case that demands to copy the records to return all the matching ones (Figure~\ref{fig:copy_p1_5k-files}), a noticeable overhead per query occurs as the engine must both locate the relevant file fragments and materialize the matching rows. In the \emph{count} case (Figure~\ref{fig:count_p1_5k-files}), execution is generally faster and more stable, because DuckDB can aggregate in-stream without returning full rows, reducing data movement and result materialization. Overall, these results show that on a single core, query latency is sensitive to file management overhead, and simple aggregations (count) are significantly cheaper than returning full records even when the logical workload is identical.

\begin{figure}
    \centering
    \begin{subfigure}[b]{\linewidth}
        \centering
        \includegraphics[width=\linewidth]{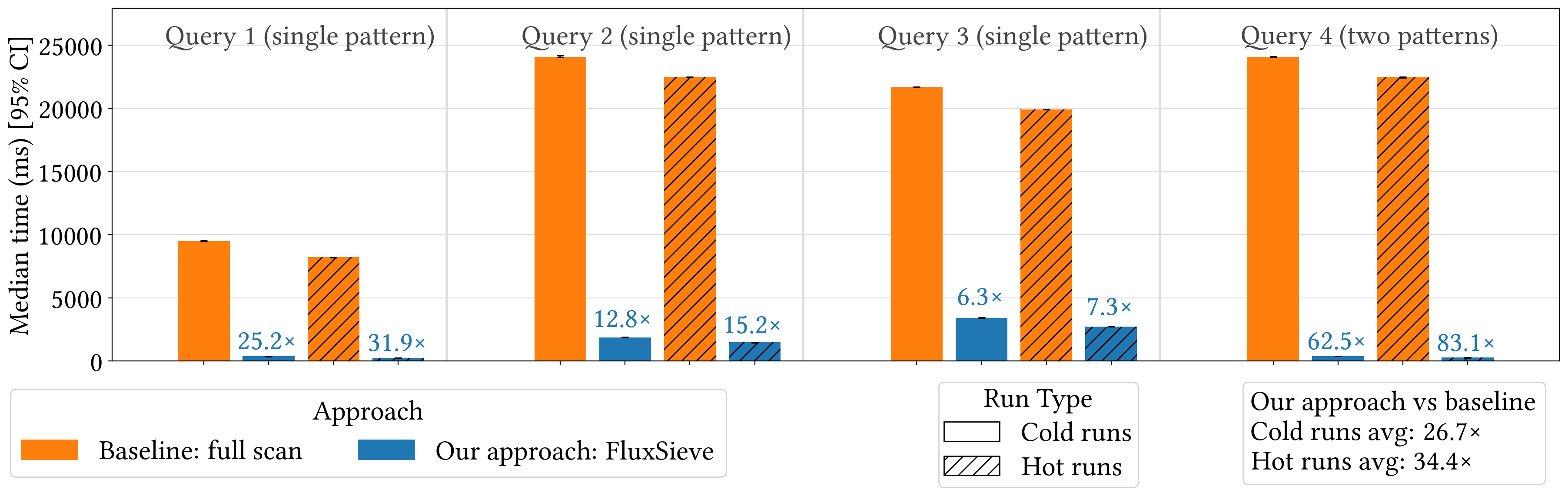}
        \caption{Returning records}
        \label{fig:copy_p1_1k-files}
    \end{subfigure}
    \vspace{0.5em}
    \begin{subfigure}[b]{\linewidth}
        \centering
        \includegraphics[width=\linewidth]{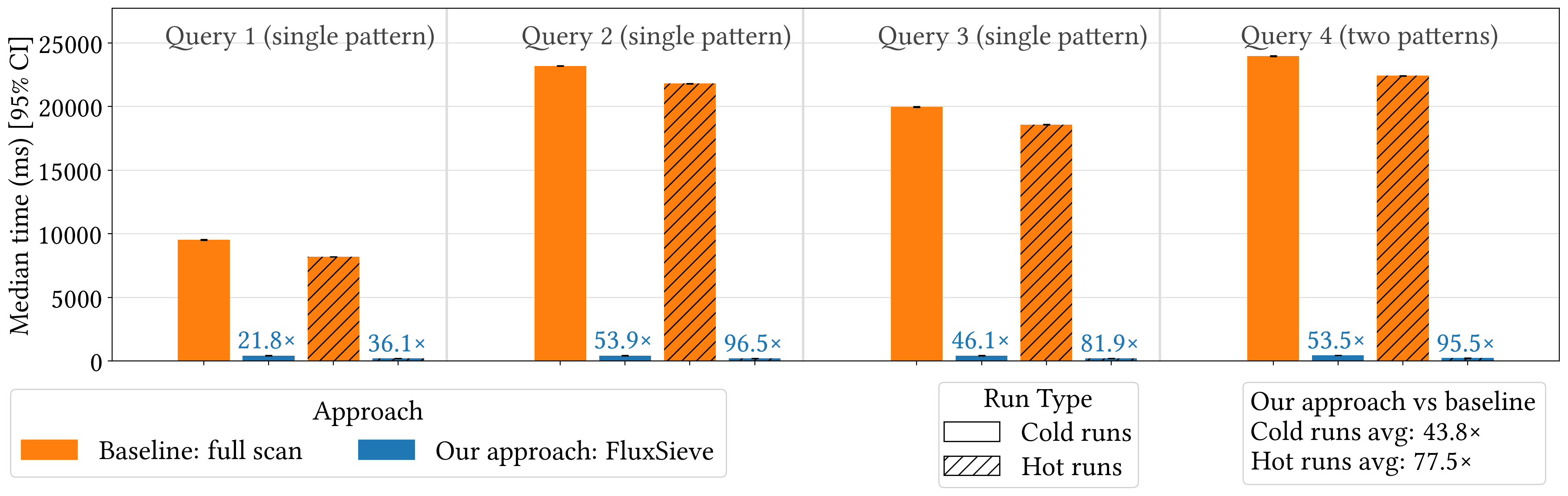}
        \caption{Counting records}
        \label{fig:count_p1_1k-files}
    \end{subfigure}
    \caption{\boldmath Parallelism level 1: $\approx$1k files, $\approx$10k records per file.}
    \label{fig:p1_1k-files}
\end{figure}

\paragraph{1 CPU core, $\approx$1k files, $\approx$10k records/file (\cref{fig:p1_1k-files}).}
Keeping the same 10 million records consolidated them into fewer, larger Parquet files reduces metadata and file-opening overhead. In the \emph{copy} scenario (Figure~\ref{fig:copy_p1_1k-files}), query times drop compared to the \cref{fig:p1_5k-files} results, reflecting more efficient sequential scanning and fewer file handles. The improvement is even more pronounced in the \emph{count} scenario (Figure~\ref{fig:count_p1_1k-files}), where DuckDB can process larger batches per file and aggregate results quickly. This highlights how file compaction (reducing the small-files problem) is beneficial even without parallelism.

\begin{figure}
    \centering
    \begin{subfigure}[b]{\linewidth}
        \centering
        \includegraphics[width=\linewidth]{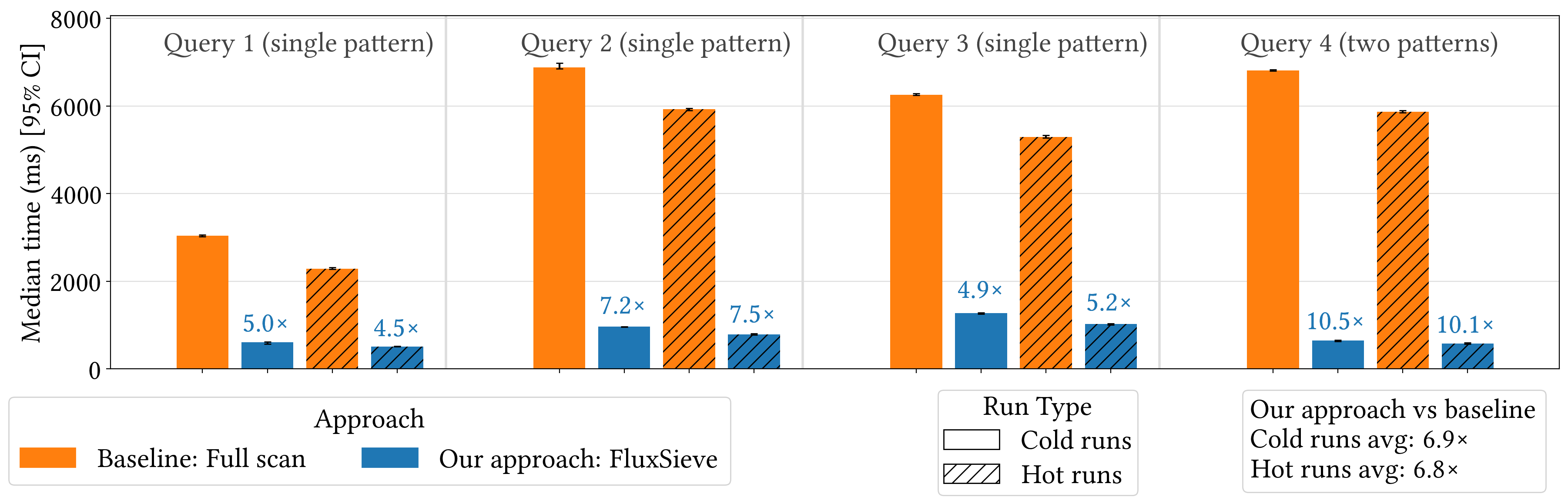}
        \caption{Returning records}
        \label{fig:copy_p4_5k-files}
    \end{subfigure}
    \vspace{0.5em}
    \begin{subfigure}[b]{\linewidth}
        \centering
        \includegraphics[width=\linewidth]{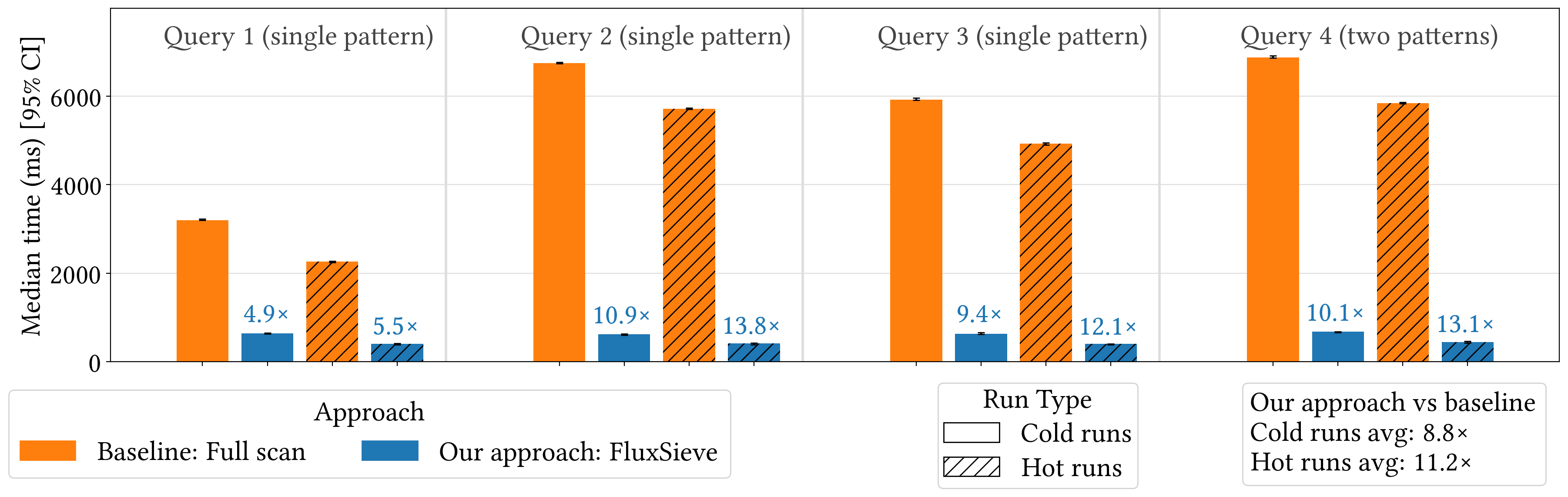}
        \caption{Counting records}
        \label{fig:count_p4_5k-files}
    \end{subfigure}
    \caption{\boldmath Parallelism level 4: $\approx$5k files, $\approx$2k records per file.}
    \label{fig:p4_5k-files}
\end{figure}

\paragraph{4 CPU cores, $\approx$5k files, $\approx$2k records/file (\cref{fig:p4_5k-files}).}
DuckDB can parallelize scanning across the files with four cores/threads. In the \emph{copy} case (Figure~\ref{fig:copy_p4_5k-files}), the parallelism compensates for part of the small-file overhead: median query times decrease compared to the 1-core / 5k-files configuration, though the gains are limited by per-file overhead and coordination costs. In the \emph{count} case (Figure~\ref{fig:count_p4_5k-files}), the combination of parallel scanning and cheap aggregation yields further performance gains, but the scaling is sublinear because the engine must coordinate many files and merge partial aggregates.

\begin{figure}
    \centering
    \begin{subfigure}[b]{\linewidth}
        \centering
        \includegraphics[width=\linewidth]{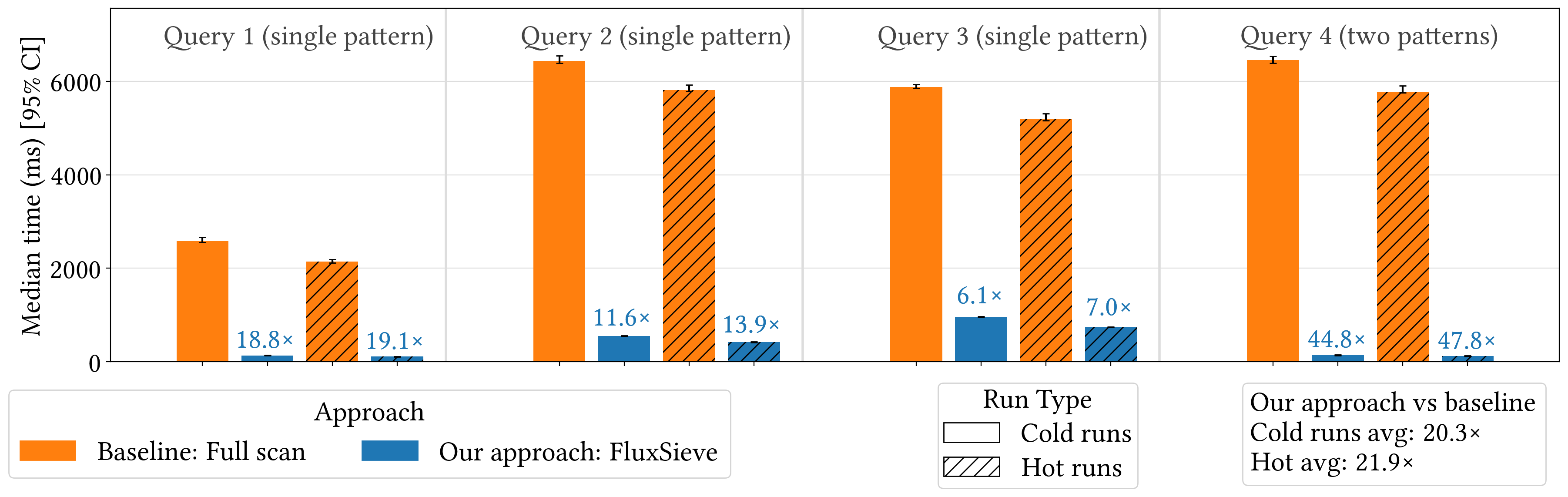}
        \caption{Returning records}
        \label{fig:copy_p4_1k-files}
    \end{subfigure}
    \vspace{0.5em}
    \begin{subfigure}[b]{\linewidth}
        \centering
        \includegraphics[width=\linewidth]{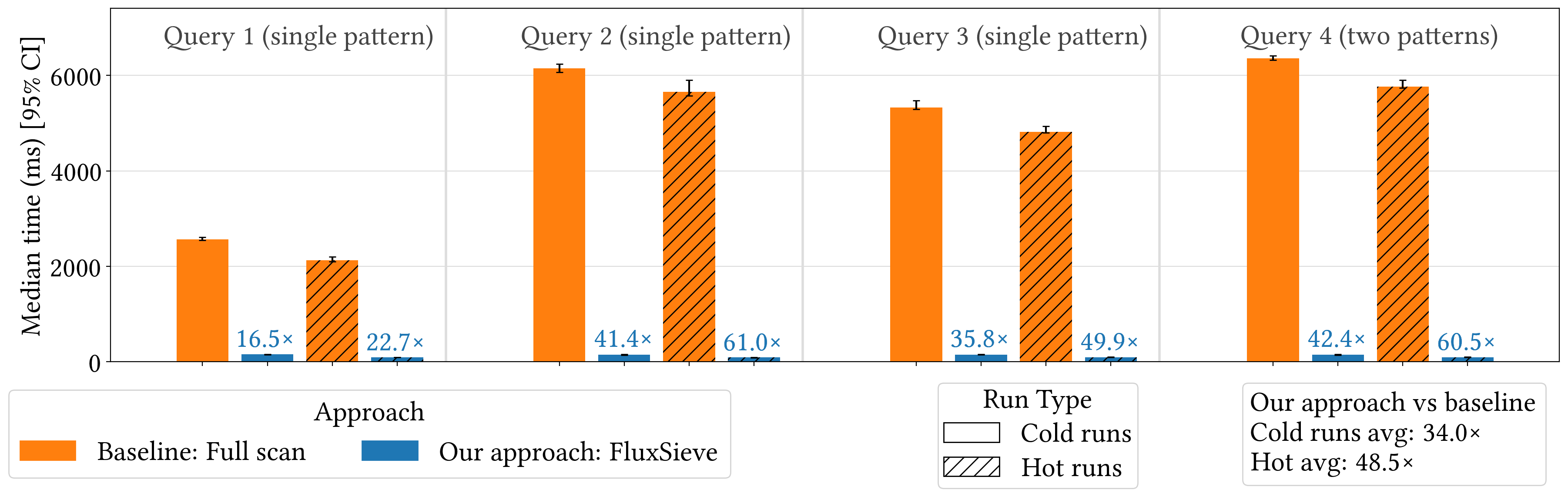}
        \caption{Counting records}
        \label{fig:count_p4_1k-files}
    \end{subfigure}
    \caption{\boldmath Parallelism level 4: $\approx$1k files, $\approx$10k records per file.}
    \label{fig:p4_1k-files}
\end{figure}

\paragraph{4 CPU cores, $\approx$1k files, $\approx$10k records/file (\cref{{fig:p4_1k-files}}).}
This is an optimal scenario with four cores and only around 1k Parquet files, DuckDB can assign larger chunks of data to each worker and amortize file-opening overhead. In the \emph{copy} scenario (Figure~\ref{fig:copy_p4_1k-files}), query latencies drop substantially, showing near-ideal use of available CPU resources. Results are even better for the \emph{count} scenario (Figure~\ref{fig:count_p4_1k-files}): queries become significantly faster, reflecting an efficient combination of parallel I/O and lightweight aggregation. 

\paragraph{Findings Summary}Across all results, four main observations emerge. First, \textbf{our approach is highly efficient, achieving its highest speedups in CPU-constrained settings}: the full-scan baseline is substantially more CPU-intensive and thus benefits more from additional cores, whereas our in-stream filtering approach is already efficient and requires less hardware to deliver low query latencies. For instance, in 1-core configurations, our in-stream filtering approach yields huge gains w.r.t the baseline: speedups for both counting and copying queries often exceed $30\times$ and can surpass $60\times$. These results indicate that, when CPU is scarce, avoiding full scans and pushing filtering into the ingestion path provides a substantial advantage. Second, \textbf{parallelism helps, but is bounded by layout}: adding CPU cores improves performance, particularly when the number of files is moderate; however, in the presence of thousands of small files, per-file overhead limits the benefits of additional cores. Finally, \textbf{counting records is faster}: although count queries involve aggregation, they avoid materializing and transferring full result rows, and are therefore faster than queries that return all matching records. 

\section{Evaluation of Analytical System}\label{sec:evaluation-olap}

Real-Time Online Analytical Processing (RTOLAP) systems are engineered to minimize query latency while ingesting high-velocity streams of records. Apache~Pinot~\cite{pinot2018} exemplifies this design, sustaining millions of events per second and serving large numbers of concurrent queries. Because our streaming enrichment layer focuses on accelerating downstream analytics, we measure potential improvements by integrating our approach into Apache Pinot's deployment. Pinot’s broker–server architecture exposes precise per-query timing, permitting statistically robust comparisons between the enriched and baseline pipelines at millisecond granularity.

It is important to measure the impact of multi-pattern matching and record enrichment on analytics data plane storage. Pinot’s columnar segments employ dictionary, run-length, and bit-packed encodings to compress column values, achieving compact on-disk footprints without sacrificing scan efficiency. Focusing the evaluation on query execution time (reported in milliseconds) aligns the measurement axes with Pinot’s core strengths: low execution time for analytics and storage-efficient columnar representation.

\subsection{Integration With Apache~Pinot}

\paragraph{Baseline.}
Application logs are produced to a Kafka partitioned topic, where partitions are assigned to Apache Pinot's server to ingest data. 
Pinot’s Kafka connector consumes each partition directly into a \texttt{REALTIME} table whose schema mirrors the original record fields. 
Relevant index configurations are:  
(1)~\emph{Default} (called \emph{Full table scan}): dictionary-encoded forward indexes;  
(2)~\emph{Default~+~Text} (called \emph{Text indexed}): the same configuration augmented with Pinot’s per-column FTS index on the text fields, \textit{json\_context}), which can accelerate regex and keyword predicates.
In our experiments, the \emph{Full table scan} configuration consistently exhibited query performance orders of magnitude slower than \emph{Text indexed}. Therefore, we omit \emph{Full table scan} from the experimental analysis, as the full-text index provides a more relevant baseline.

\paragraph{FluxSieve with in-stream filtering and enrichment.}
It includes a stream processor (\cref{sec:matching,sec:adaptation}) between Kafka and Pinot. 
For every log record, the processor applies 1\,000 Boolean filtering rules using multi-pattern matching. 
\emph{Enriched} record extending the baseline schema with 1\,000 additional Boolean columns \texttt{rule\_1} \dots \texttt{rule\_\!1000} are emitted.\footnote{We materialize additional Boolean columns to simplify the analysis. More efficient representations could be used in production, e.g., storing the matches in a sparse list.} 
Exactly one field is set to \texttt{TRUE} for each match; all others default to \texttt{FALSE}. 
The enriched stream is published to another Kafka topic\footnote{The additional Kafka topic is used to provide a reliable way to evaluate our approach. This additional topic could be avoided by moving the records directly from the stream processor to Apache Pinot using connectors.} where Pinot ingests the records into a table similar to the \emph{Default~+~Text} (large text columns with FTS indexes).\footnote{It is important to note that the queries executed over our approach’s table query the enriched fields. However, FTS indexes are still created in our approach to ensure a fair comparison as less intensive queries can still simply query over FTS indexes.} 

We employ one Pattern Matching Engine instance per record text field to be filtered (five in total), all running within the same stream processor. Each instance evaluates its corresponding field and filtering rules. Records that match any of these rules are enriched with the identifier. This design isolates the cost of in-stream filtering/enrichment while holding Pinot’s indexing strategy. 

\subsection{Setup}

Apache Pinot was deployed on a Kubernetes cluster backed by Amazon EBS \texttt{gp3} volumes formatted with the \texttt{ext4} filesystem. The Pinot cluster follows the standard controller--broker--server architecture: a single controller, a single broker, and four server instances to provide a reasonable degree of query parallelism. Pinot controller and broker pods, as well as each Pinot server pod, are scheduled on \texttt{m6i.xlarge} instances. Each Pinot server pod runs on a dedicated instance, while the other pods share one instance.\footnote{For an in-depth description of Apache Pinot's architecture and deployment, we refer interested readers to the original system paper~\cite{pinot2018} and to its documentation~\cite{pinot2026}.}

For data ingestion, we use two Apache Kafka brokers deployed on Kubernetes using Strimzi, each running on \texttt{m6i.2xlarge} instances. The input data is written to Kafka topics with 40 partitions, which enables parallel ingestion into Pinot servers and supports high-throughput for real-time analytical workloads.

\subsection{Query Performance}

In this experimental scenario, \Cref{sec:uhsq} describes the \emph{Ultra-high Selective Queries} scenario, where matching records are extremely rare, which aligns with the context and motivation of this work (see \cref{sec:background}). In addition, query performance is also evaluated in a second scenario, \emph{High Selective Queries} (\Cref{sec:hsq}), where matches are less rare and, when present, occur in quantities an order of magnitude higher than in \Cref{sec:uhsq}. These scenarios with different levels of selectivity were chosen to assess whether similar performance trends hold under varying record matching. Moreover, for cold queries, whose execution requires infrastructure overhead due to environment redeployment, we repeated each query execution 10 times. As for hot queries, no redeployment is needed, we repeated each query 40 times.

\subsubsection{Ultra-high Selective Queries}\label{sec:uhsq}

\begin{figure}
    \centering
    \includegraphics[width=\linewidth]{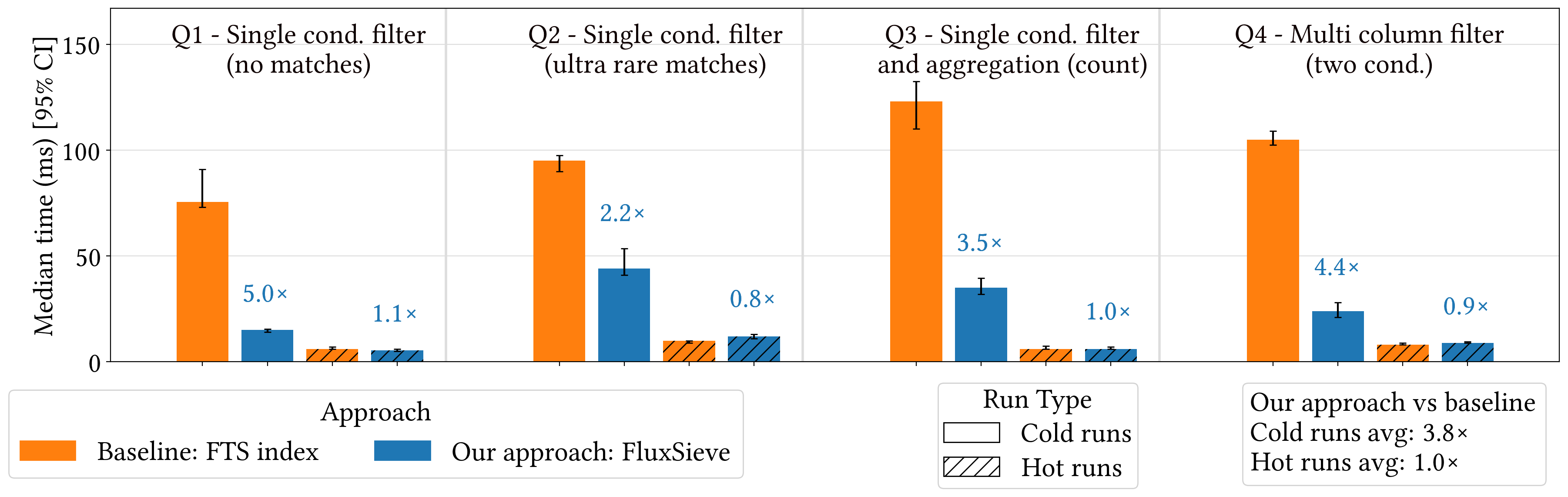}
    \caption{5 million records.}
    \label{fig:5m}
\end{figure}

\paragraph{5 Million records (Figure~\ref{fig:5m}).}

For the dataset with 5 million (5M) records, the median query execution time clearly differs across the evaluated approaches: text-indexed search and our in-stream filtering technique. Text-indexed search can improve query performance substantially but still incurs noticeable overhead for ultra-highly selective queries. In contrast, our in-stream filtering approach achieves the lowest median execution times for almost all queries, both on cold and hot data. The gap between our method and the text-indexed baseline is particularly notable in the scenario of runs on cold data, which is mostly due to the data pruning possible with our approach that avoids I/O bottlenecks. This indicates that, even at a moderate dataset size, exploiting in-stream filtering yields tangible performance gains under ultra-high selectivity.

\begin{figure}
    \centering
    \includegraphics[width=\linewidth]{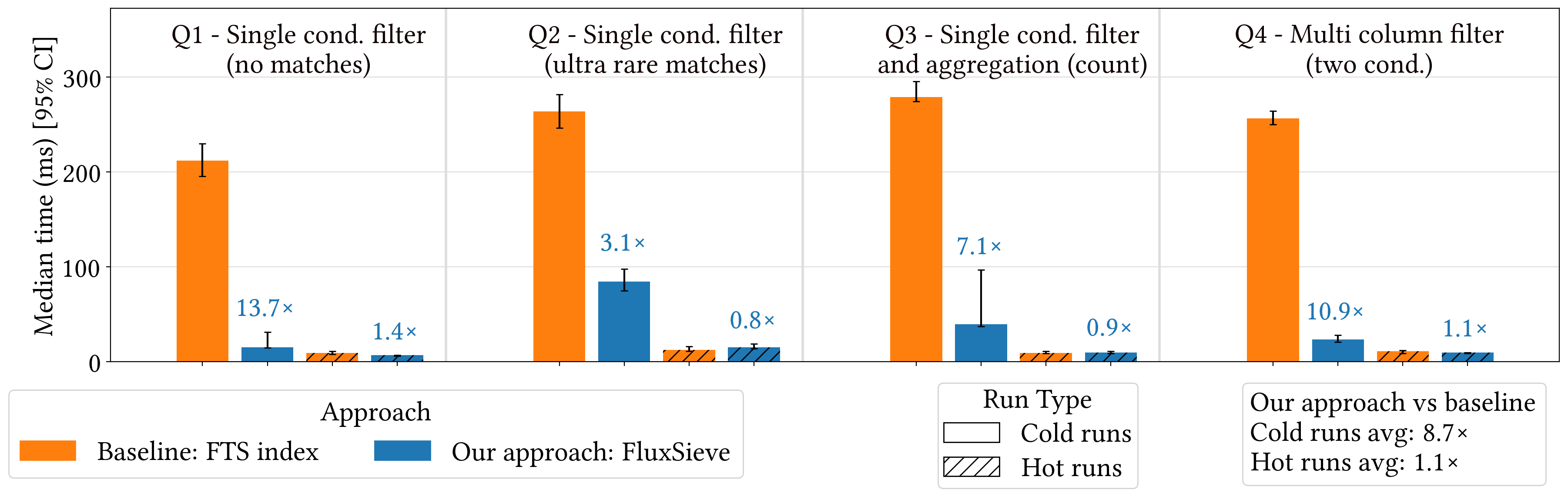}
    \caption{10 million records.}
    \label{fig:10m}
\end{figure}

\paragraph{10 million records (Figure~\ref{fig:10m}).}
With 10 million (10M) records, the approaches exhibit higher absolute execution times due to the larger dataset, yet the relative ordering between the methods remains similar. The text-indexed baseline is consistently slower than our in-stream filtering approach for almost all queries, both on cold and hot data. While the increase in query time from 5 to 10 million records is observable, the growth for our approach is less pronounced than for the text-indexed baseline, showing better scalability and speedups in our approach as data volume increases. 

\begin{figure}
    \centering
    \includegraphics[width=\linewidth]{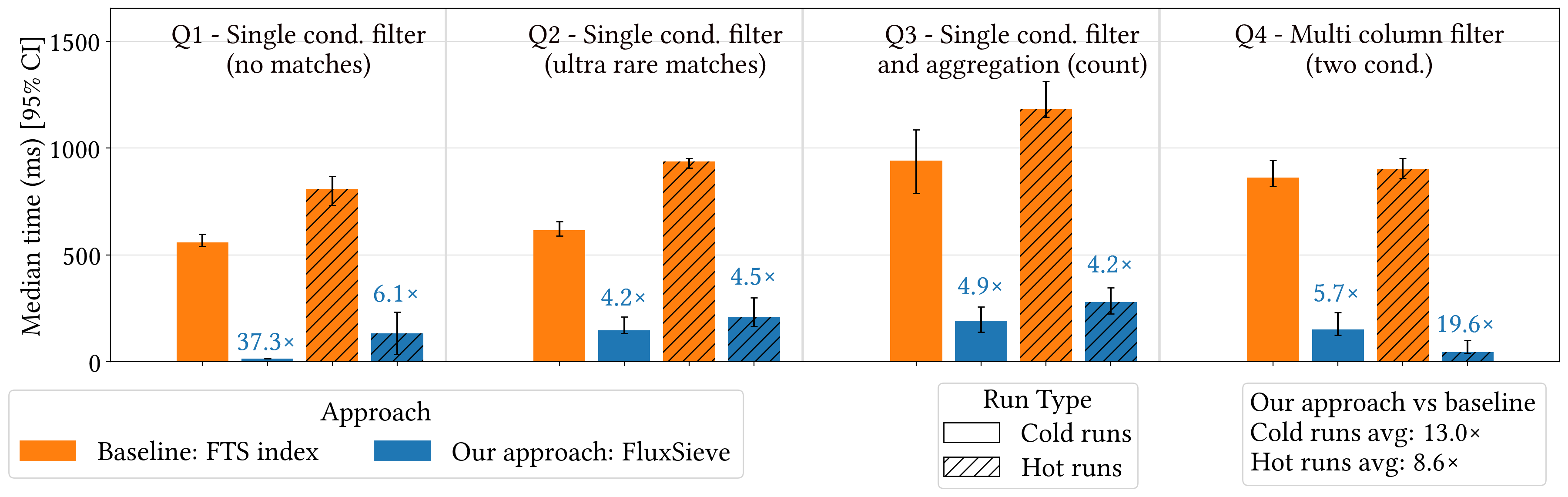}
    \caption{20 million records.}
    \label{fig:20m}
\end{figure}

\paragraph{20 million records (Figure~\ref{fig:20m}).}
At 20 million (20M) records, the separation between the two main approaches becomes more evident. The median execution times for text-indexed queries increase substantially, while our in-stream filtering approach maintains consistently lower query times across the board, again for both cold and hot runs. The fact that our method remains clearly outperforming the text-indexed baseline for all queries at this scale highlights that the benefits of in-stream filtering amplify as the dataset grows under ultra-high selectivity. The relation between cold and hot data runs also indicates that the approach is robust to cache state, which is important in deployment scenarios where queries may frequently hit cold data. The plots also illustrate that the performance advantage of in-stream filtering is preserved regardless of whether data is served from cold or hot caches, reinforcing that the gains stem from execution strategy rather than from caching effects alone.

\begin{figure}
    \centering
    \includegraphics[width=\linewidth]{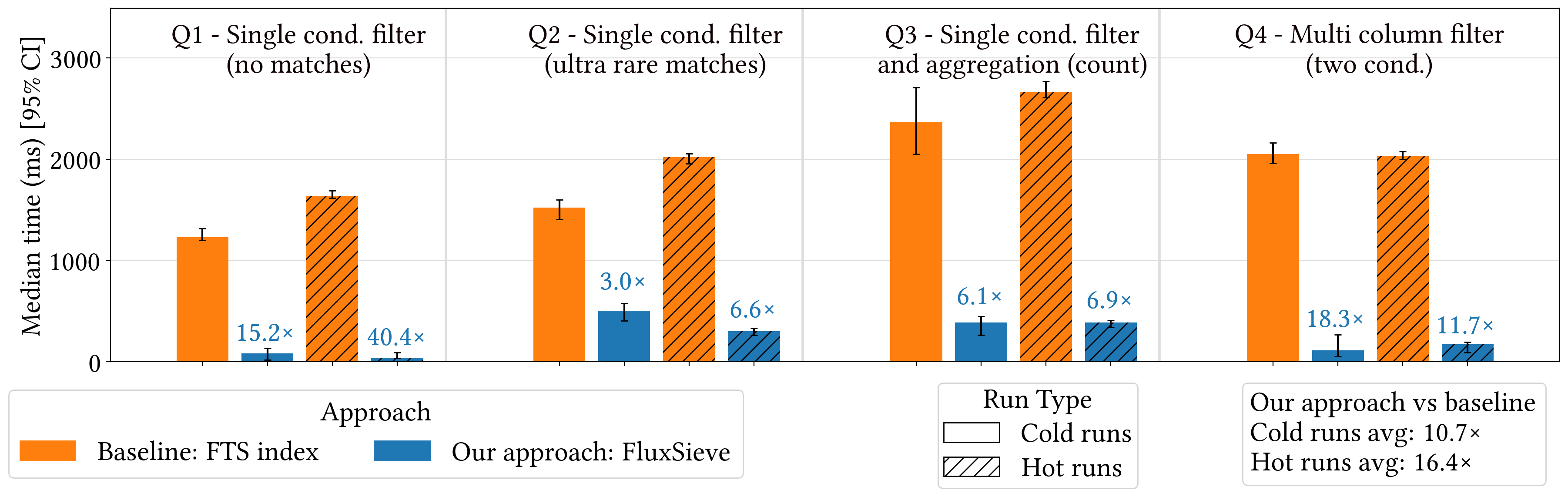}
    \caption{40 million records.}
    \label{fig:40m}
\end{figure}

\paragraph{40 million records (Figure~\ref{fig:40m}).}
When scaling to 40 million (40M) records, the trends observed at smaller scales persist and become even more pronounced. The text-indexed baseline shows noticeably higher median execution times, reflecting the cost of indexing structures and lookups at this size. Our FluxSieve approach delivers significantly lower execution time for all queries, on both cold and hot data. The parallel progression of the curves for the two approaches suggests that, in this ultra-high selectivity scenario and as the dataset grows, our method maintains and increases its gains. This indicates that FluxSieve can handle large-scale deployments while preserving predictable and low query execution times.

\begin{figure}
    \centering
    \includegraphics[width=\linewidth]{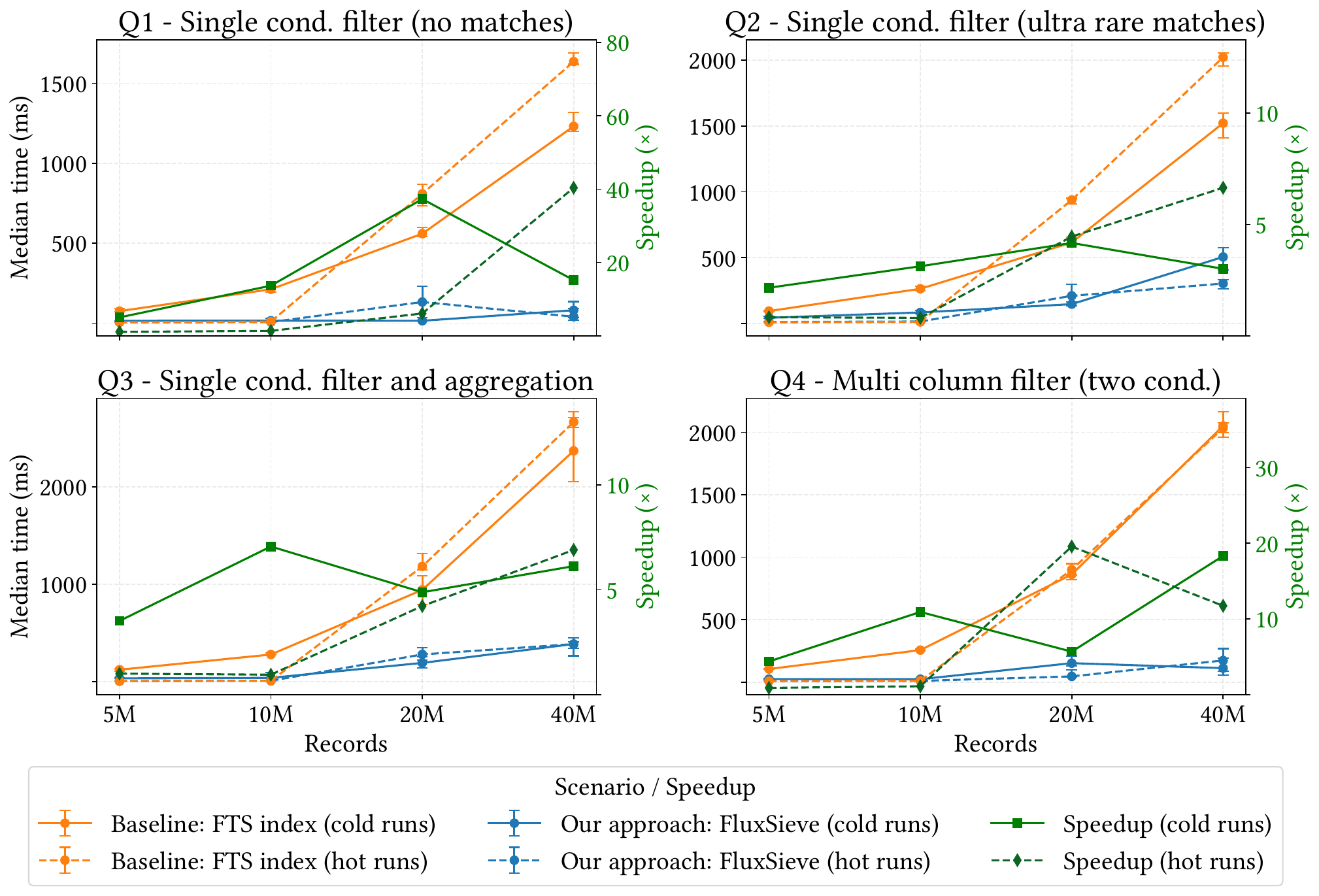}
    \caption{Ultra-High Selectivity: Overall comparison.}
    \label{fig:ultra-overall}
\end{figure}

\paragraph{Overall speedup versus text-indexed search (Figure~\ref{fig:ultra-overall}).}
The aggregated view in Figure~\ref{fig:ultra-overall} summarizes the speedup of our in-stream filtering approach over the text-indexed baseline across all dataset sizes and query types. For all query variants (pure filter, filter plus aggregation, and multi-filter), our method achieves performance gains across both cold and hot data. The speedup values are generally greater than one and often reach several times faster execution. As the dataset size increases from 5 to 40 million records, the speedups tend to increase, particularly for cold data runs, indicating that the relative cost of the text-indexed baseline grows faster than that of in-stream filtering. These results confirm that our approach systematically outperforms text-indexed search and that these advantages scale with data size and query complexity.

\subsubsection{High Selectivity: Extended Scenarios with Counting Aggregations}\label{sec:hsq}

\begin{figure}
    \centering
    \includegraphics[width=1.04\linewidth]{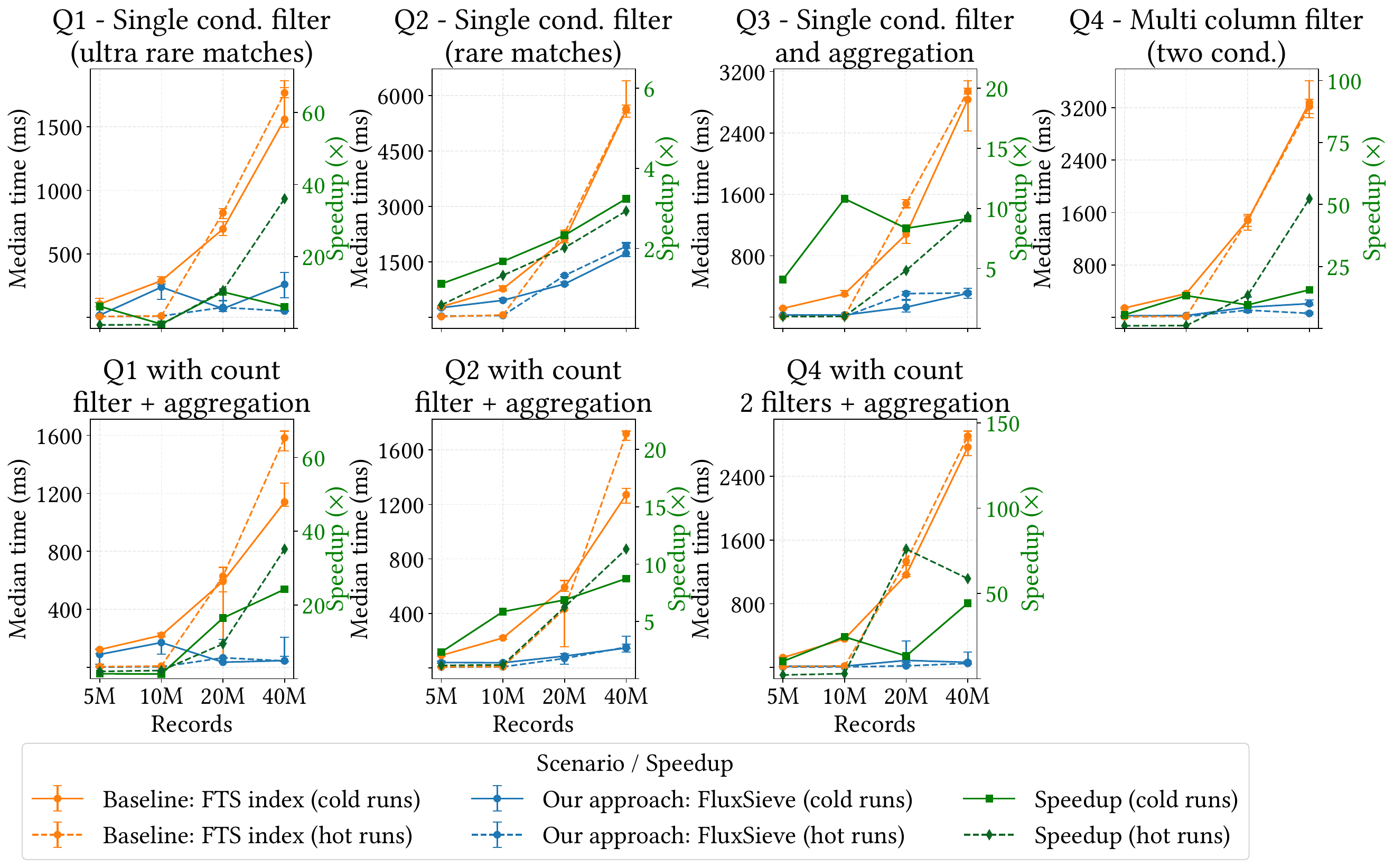}
    \caption{High Selectivity: Overall comparison.}
    \label{fig:high-overall}
\end{figure}

Figure~\ref{fig:high-overall} presents the performance of our in-stream filtering approach compared to text-indexed search in a \emph{high selectivity} scenario, i.e., where matching records are still relatively rare but notably more frequent than in the ultra-high selectivity setting. In addition to the four original query types, this figure introduces extended scenarios for Queries~1, 2, and 4, where the system not only filters records but also computes a \emph{count} of the matches (i.e., a simple aggregation). These extended queries are denoted as:
\begin{itemize}
  \item \emph{Query 1 with count}: Filter + aggregation (count),
  \item \emph{Query 2 with count}: Filter + aggregation (count),
  \item \emph{Query 4 with count}: Two filters + aggregation (count).
\end{itemize}

These queries correspond to scenarios the full records are not needed, but only a summary of the number of matches. The introduction of these queries is important because for many real-world applications (e.g., dashboards, alerts) can issue queries that aggregate over a filtered subset of data rather than demanding the full raw records. Moreover, aggregation alters the execution profile: when only a count is required, the engine can avoid materializing and transferring all matching rows, reducing memory and network pressure. Evaluating Queries~1, 2, and 4 with and without counting measures whether the performance advantages of FluxSieve persists under different query semantics.

\paragraph{Comparison with high selectivity filter-only queries.}
Across queries such as~1, 2, and 4, the results show that the performance of ``with count'' variants exhibits similar patterns of the pure filter versions. For each dataset size, the FluxSieve approach remains consistently faster than the text-indexed baseline, and the speedup curves for the ``with count'' queries are sometimes even higher than those observed with filter-only. This indicates that adding aggregations does not erode the benefits of in-stream filtering. 

From the perspective of speedup, both cold and hot data runs show that our approach maintains speedups for all query variants, with the more complex queries (e.g., Query~4 and Query~4 with count) often exhibiting the highest speedups. This suggests that as queries become more selective and/or more complex (multiple filters plus aggregation), the relative cost of the text-indexed baseline grows faster than that of in-stream filtering, even when the operation requested is a count rather than full record retrieval.

\paragraph{Findings Summary} Our approach under different selectivity, data sizes, and query types shows consistent performance gains:\footnote{These performance gains come at a negligible cost in terms of storage. In our experiments with Apache Pinot’s default compression and encoding configurations, the data size difference between our approach and the baseline was consistently below 2\%.}
\begin{itemize}
  \item Although under less extreme selective queries and aggregations the absolute execution times increase, the speedups of our approach, FluxSieve, remain significant. This confirms that FluxSieve's benefits are not restricted to extreme ultra-high selectivity.
  \item The scaling trends with dataset size (from 5 to 40 million records) observed under ultra-high selectivity carry over to the high selectivity setting. FluxSieve scales better than baseline FTS index as data volume grows. This indicates that FluxSieve is scalable.
  \item For queries over hot data, our approach provides significant speedups when the dataset does not fit in main memory (e.g., more than 20 million records). For cold data queries, our approach yields substantial speedups across all scenarios. The contrast between cold and hot runs highlights the impact of FluxSieve's benefits and optimizations achieved such as data pruning.
    
\end{itemize}

\section{Related Work}\label{sec:related}

Considering the emergence of streaming databases as systems that continuously ingest data, serve low‑latency queries on live streams, and seamlessly support queries over historical data~\cite{dulay2025}, our approach can be viewed as a refined streaming database that unifies the streaming and analytical data planes. FluxSieve focuses particularly on optimizing complex filtering with multi-pattern matching and unifying it with the analytical data plane. Key related techniques are \emph{materialized views}~\cite{mami2012} and \emph{incremental view maintenance (IVM)}~\cite{Olteanu2024,budiu2024}. Materialized views precompute and store query results, while IVM updates them incrementally as data changes. These mechanisms can improve query performance but introduce overheads in complexity, storage footprint, and adaptation to dynamic records data schema~\cite{wingerath2020}.

\citet{Winter2020} propose a hybrid view maintenance strategy for streaming data that splits work between insert and query time. Instead of fully materializing stream tuples at insertion or deferring all processing to query time, their system processes each tuple only up to the first materialization point, and completes the remaining query processing when results are requested. Implemented in the Umbra compiled query engine, this ``split maintenance'' approach achieves insert throughput competitive with dedicated stream processors while maintaining low query latency. Their approach outperformed Apache Flink on analytical workloads and traditional IVM approaches, while remaining tightly integrated with the database to combine real-time and historical data.

\citet{zhang2025} focus on real-time data processing engine on Alibaba Cloud's AnalyticDB that addresses limitations of Lambda ETL architectures and IVM. Their system combines IVM with Zero-ETL to support high-throughput, streaming ETL, and preserving data consistency, removing the need for stream processing engines.

Many works are complementary to our approach focusing on accelerating pattern matching in databases~\cite{Wang2025,reichinger2024}. \citet{Korber21} propose an index method for filtering queries on event stores to perform data prefiltering. However, their design remains limited in scalability because it relies on sequential pattern matching rather than multi-pattern matching and does not support processing over multiple partitions. Furthermore, the integration between the streaming and database planes is limited. In contrast, to the best of our knowledge, our proposal provides a more integrated architectural design and processing model, combining in-stream, multi-pattern filtering with tight coupling to analytical systems.

\section{Closing Remarks}\label{sec:closing}

This paper explored how to reconcile push-based stream processing with pull-based analytical querying in large-scale observability platforms by \emph{moving expensive, repetitive queries out of the analytical plane and into the ingestion pipeline}. Rather than relying solely on table scans and heavyweight text indexes at query time, we present FluxSieve which embeds an efficient in-stream precomputation layer, while keeping the analytical plane as the source of truth.

\paragraph{Key Findings and Implications}
Across different systems:

\begin{itemize}
  \item Query performance improves up to orders of magnitude.
  \item Ingestion throughput's overhead is not high as well as CPU usage overhead is manageable for the trade-off with the benefits of our approach.
  \item Storage overhead can be negligible.
  \item Benefits generalize across query types, analytical systems, configurations, data volumes, and deployment modes.
\end{itemize}

These results indicate that, for observability-style workloads with intensive, and often ultra-selective queries, shifting filtering logic into the streaming data plane is a powerful design pattern. It reduces the work required at query time and allows analytical engines to operate on data that is stored in optimal shape for retrieval. These findings point toward viable approaches beyond traditional full-text search (FTS) indexes: expensive and repetitive queries instead of FTS index can rely on in-stream precomputation, while less intensive or ad-hoc queries can fall back on lighterweight indexing strategies or even full-table scans when appropriate. This tiered approach enables efficient allocation of system resources, reserving heavyweight indexing for scenarios where it provides true benefits rather than applying it uniformly across all query patterns.

More broadly, our findings suggest that ``turning the database inside out''~\cite{kleppmann2017} for selected, highly selective queries is not only conceptually appealing, but also empirically effective when combined with modern multi-pattern matching engines and columnar storage. Furthermore, the proposed approach is not limited to observability: it can be applied more generally to other domains with repetitive analytical queries that benefit from a unified architecture, such as generic repetitive queries~\cite{Feng2011} and updating continuous queries~\cite{Bonifati2024}.

\paragraph{Limitations}

This work is deliberately specific: (I) We target \textbf{highly selective filtering conditions}. (II) Our evaluation uses \textbf{log-like datasets} with controlled schemas and query patterns. While they are designed to be representative, other scenarios may exhibit additional complexities. (III) We focus on \textbf{pattern-based filtering and aggregations}. Other in-stream computations (e.g., windowed joins) may exhibit different trade-offs.
These limitations are not fundamental, but they delineate the scope in which our current results should be interpreted.

\paragraph{Future Work}
Our approach opens several promising directions: (I) \textbf{Beyond filtering:} Many workloads could benefit from in-stream aggregations, studying the right balance between precomputation and flexibility is an important next step. (II) Deeper \textbf{integration with open table formats}: integrating our approach with open formats could further increase its generalization. (III) \textbf{Newer pattern matching engines}: new approaches such as BLARE~\cite{zhang2023} may further reduce the CPU cost of in-stream filtering.

\balance
\bibliographystyle{ACM-Reference-Format}
\bibliography{references}

@book{kleppmann2017,
  title     = {{Designing Data-Intensive Applications: The Big Ideas Behind Reliable, Scalable, and Maintainable Systems}},
  author    = {Kleppmann, Martin},
  year      = {2017},
  publisher = {O'Reilly Media, Inc.},
  address   = {Sebastopol, CA},
  isbn      = {978-1-4493-7332-0},
  pages     = {616}
}

@article{Purtzel2025,
author = {Purtzel, Steven and Weidlich, Matthias},
title = {SuSe: Summary Selection for Regular Expression Subsequence Aggregation over Streams},
year = {2025},
issue_date = {June 2025},
publisher = {Association for Computing Machinery},
address = {New York, NY, USA},
volume = {3},
number = {3},
url = {https://doi.org/10.1145/3725359},
doi = {10.1145/3725359},
journal = {Proceedings of the 2025 International Conference on Management of Data},
month = jun,
articleno = {222},
numpages = {27}
}

@article{budiu2024,
  title={DBSP: Incremental computation on streams and its applications to databases},
  author={Budiu, Mihai and Chajed, Tej and McSherry, Frank and Ryzhyk, Leonid and Tannen, Val},
  journal={ACM SIGMOD Record},
  volume={53},
  number={1},
  pages={87-95},
  year={2024},
  publisher={ACM New York, NY, USA}
}

@inproceedings{pinot2018,
  title={{Pinot: Realtime OLAP for 530 Million Users}},
  author={Im, Jean-Fran{\c{c}}ois and Gopalakrishna, Kishore and Subramaniam, Subbu and Shrivastava, Mayank and Tumbde, Adwait and Jiang, Xiaotian and Dai, Jennifer and Lee, Seunghyun and Pawar, Neha and Li, Jialiang and others},
  booktitle={Proceedings of the 2018 International Conference on Management of Data},
  pages={583-594},
  year={2018},
  publisher={ACM},
  address={Houston, TX, USA}
}

@article{Wang2025,
  author    = {Wang, Ziheng and Wei, Junyu and Aiken, Alex and Zhang, Guangyan and T{\o}rring, Jacob O and Jiang, Rain and Jiang, Chenyu and Xu, Wei},
  title     = {LogCloud: Fast Search of Compressed Logs on Object Storage},
  journal   = {Proceedings of the VLDB Endowment},
  volume    = {18},
  number    = {8},
  pages={2362--2370},
  year      = {2025}
}

@misc{reichinger2024,
      title={COPR -- Efficient, large-scale log storage and retrieval}, 
      author={Julian Reichinger and Thomas Krismayer and Jan Rellermeyer},
      year={2024},
      eprint={2402.18355},
      archivePrefix={arXiv},
      primaryClass={cs.IR},
      url={https://arxiv.org/abs/2402.18355}, 
}

@article{Grandi2022,
title = {Unleashing the power of querying streaming data in a temporal database world: A relational algebra approach},
journal = {Information Systems},
volume = {103},
pages = {101872},
year = {2022},
issn = {0306-4379},
doi = {https://doi.org/10.1016/j.is.2021.101872},
author = {Fabio Grandi and Federica Mandreoli and Riccardo Martoglia and Wilma Penzo}
}

@inproceedings{Bonifati2024,
author = {Bonifati, Angela and Tommasini, Riccardo},
title = {An Overview of Continuous Querying in (Modern) Data Systems},
year = {2024},
isbn = {9798400704222},
publisher = {Association for Computing Machinery},
address = {New York, NY, USA},
url = {https://doi.org/10.1145/3626246.3654679},
doi = {10.1145/3626246.3654679},
booktitle = {Companion of the 2024 International Conference on Management of Data},
pages = {605–612},
numpages = {8},
location = {Santiago, Chile}
}

@misc{pinot2026,
  title        = {Apache Pinot -- Official Website},
  author = {Apache Pinot},
  howpublished = {\url{https://pinot.apache.org/}},
  note         = {Accessed: 2026-03-02},
  organization = {Apache Software Foundation},
  year         = {2026}
}

@inproceedings{wingerath2020,
  title     = {InvaliDB: scalable push-based real-time queries on top of pull-based databases},
  author    = {Wingerath, Wolfram and Gessert, Felix and Ritter, Norbert},
  booktitle = {2020 IEEE 36th International Conference on Data Engineering (ICDE)},
  pages     = {1874--1877},
  year      = {2020},
  publisher = {IEEE},
  address   = {Piscataway, NJ, USA}
}

@inproceedings{Feng2011,
author = {Yu, Feng and Hou, Wen-Chi and Luo, Cheng and Zhu, Qiang and Che, Dunren},
title = {Join selectivity re-estimation for repetitive queries in databases},
year = {2011},
isbn = {9783642230905},
publisher = {Springer},
address = {Berlin},
booktitle = {Proceedings of the 22nd International Conference on Database and Expert Systems Applications},
pages = {420–427},
numpages = {8},
location = {Toulouse, France},
series = {DEXA'11}
}

@article{Winter2020,
  title={Meet Me Halfway: Split Maintenance of Continuous Views},
  author={Winter, Christian and Schmidt, Tobias and Neumann, Thomas and Kemper, Alfons},
  journal={Proceedings of the VLDB Endowment},
  volume={13},
  number={11},
  pages={2620--2633},
  year={2020},
  doi={10.14778/3407790.3407849}
}

@article{vogel2022,
  title={Self-adaptation on parallel stream processing: A systematic review},
  author={Vogel, Adriano and Griebler, Dalvan and Danelutto, Marco and Fernandes, Luiz Gustavo},
  journal={Concurrency and Computation: Practice and Experience},
  volume={34},
  number={6},
  pages={e6759},
  year={2022},
  publisher={Wiley Online Library}
}

@article{abadi2020,
  title={The Seattle report on database research},
  author={Abadi, Daniel and Ailamaki, Anastasia and Andersen, David and Bailis, Peter and Balazinska, Magdalena and Bernstein, Philip and Boncz, Peter and Chaudhuri, Surajit and Cheung, Alvin and Doan, AnHai and others},
  journal={ACM Sigmod Record},
  volume={48},
  number={4},
  pages={44-53},
  year={2020},
  publisher={ACM New York, NY, USA}
}

@book{dulay2025,
  author    = {Hubert Dulay and Ralph Matthias Debusmann},
  title     = {Streaming Databases},
  publisher = {O'Reilly Media, Inc.},
  year      = {2025},
  address   = {Sebastopol, CA},
  isbn      = {9781098154820},
  url       = {https://www.oreilly.com/library/view/streaming-databases/9781098154820/}
}

@article{Fragkoulis2023,
    author = {Fragkoulis, Marios and Carbone, Paris and Kalavri, Vasiliki and Katsifodimos, Asterios},
    title = {A survey on the evolution of stream processing systems},
    year = {2024},
    volume = {33},
    number = {2},
    doi = {10.1007/s00778-023-00819-8},
    journal = {The VLDB Journal},
    pages = {507–541},
    publisher={Springer}
}

@inproceedings{armbrust2021,
  author       = {Matei Zaharia and
                  Ali Ghodsi and
                  Reynold Xin and
                  Michael Armbrust},
  title        = {Lakehouse: {A} New Generation of Open Platforms that Unify Data Warehousing
                  and Advanced Analytics},
  booktitle    = {11th Conference on Innovative Data Systems Research, {CIDR} 2021,
                  Virtual Event, January 11-15, 2021, Online Proceedings},
  publisher    = {www.cidrdb.org},
  year         = {2021},
  address      = {Virtual Event},
  pages        = {1-8}
}

@inproceedings{SEAA2023,
    author = {Vogel, Adriano and Henning, S\"{o}ren and Ertl, Otmar and Rabiser, Rick},
    title = {A systematic mapping of performance in distributed stream processing systems},
    booktitle = {Euromicro Conference on Software Engineering and Advanced Applications},
    year = {2023},
    month = sep,
    doi = {10.1109/SEAA60479.2023.00052},
    publisher={IEEE},
      pages={293-300},
    address = {Durres, Albania},
}

@article{zhang2023,
  title={Exploiting Structure in Regular Expression Queries},
  author={Zhang, Ling and Deep, Shaleen and Floratou, Avrilia and Gruenheid, Anja and Patel, Jignesh M and Zhu, Yiwen},
  journal={Proceedings of the ACM on Management of Data},
  volume={1},
  number={2},
  pages={1-28},
  year={2023},
  publisher={ACM New York, NY, USA}
}

@article{giouroukis2025,
  title={Analyzing Near-Network Hardware Acceleration with Co-Processing on DPUs},
  author={Giouroukis, Dimitrios and Nugroho, Dwi PA and Pandey, Varun and Zeuch, Steffen and Markl, Volker},
  journal={Proceedings of the VLDB Endowment},
  volume={18},
  number={13},
  pages={5689-5702},
  year={2025},
  publisher={VLDB Endowment}
}

@inproceedings{raasveldt2018,
  title={Fair benchmarking considered difficult: Common pitfalls in database performance testing},
  author={Raasveldt, Mark and Holanda, Pedro and Gubner, Tim and M{\"u}hleisen, Hannes},
  booktitle={Proceedings of the Workshop on Testing Database Systems},
  pages={1-6},
  year={2018},
  publisher={ACM},
  address={Houston, TX, USA}
}

@inproceedings{hoefler2015,
  title={Scientific benchmarking of parallel computing systems: twelve ways to tell the masses when reporting performance results},
  author={Hoefler, Torsten and Belli, Roberto},
  booktitle={Proceedings of the international conference for high performance computing, networking, storage and analysis},
  pages={1-12},
  year={2015},
  publisher={ACM},
  address={Austin, TX, USA}
}

@inproceedings{ICPE2024,
    author = {Henning, S\"{o}ren and Vogel, Adriano and Leichtfried, Michael and Ertl, Otmar and Rabiser, Rick},
    title = {ShuffleBench: A Benchmark for Large-Scale Data Shuffling Operations with Distributed Stream Processing Frameworks},
    year = {2024},
    isbn = {9798400704444},
    publisher = {ACM},
    address = {New York, NY, USA},
    doi = {10.1145/3629526.3645036},
    booktitle = {ACM/SPEC International Conference on Performance Engineering},
    pages = {2-13},
    numpages = {12}
}

@inproceedings{DEBS2024,
    author = {Vogel, Adriano and Henning, S\"{o}ren and Perez-Wohlfeil, Esteban and Ertl, Otmar and Rabiser, Rick},
    title = {A Comprehensive Benchmarking Analysis of Fault Recovery in Stream Processing Frameworks},
    year = {2024},
    isbn = {9798400704437},
    publisher = {ACM},
    address = {New York, NY, USA},
    doi = {10.1145/3629104.3666040},
    booktitle = {Proceedings of the 18th ACM International Conference on Distributed and Event-Based Systems},
    pages = {171-182},
    numpages = {12},
    location = {Villeurbanne, France},
    series = {DEBS '24}
}

@article{aho1975,
  title={Efficient string matching: an aid to bibliographic search},
  author={Aho, Alfred V and Corasick, Margaret J},
  journal={Communications of the ACM},
  volume={18},
  number={6},
  pages={333-340},
  year={1975},
  publisher={ACM New York, NY, USA}
}

@article{zhang2025,
  title={Streaming View: An Efficient Data Processing Engine for Modern Real\-time Data Warehouse of Alibaba Cloud},
  author={Zhang, Fangyuan and Wu, Mengqi and Xu, Chunlei and Bao, Yunong and Qiao, Jiyu and Zhou, Yingli and Fan, Hua and Yin, Caihua and Zhou, Wenchao and Li, Feifei},
  journal={Proceedings of the VLDB Endowment},
  volume={18},
  number={12},
  pages={5153-5165},
  year={2025},
  publisher={VLDB Endowment}
}

@inproceedings{raasveldt2019,
  title={{Duckdb: an embeddable analytical database}},
  author={Raasveldt, Mark and M{\"u}hleisen, Hannes},
  booktitle={Proceedings of the 2019 international conference on management of data},
  pages={1981-1984},
  year={2019},
  publisher={ACM},
  address={Amsterdam, Netherlands}
}

@inproceedings{Olteanu2024,
author = {Olteanu, Dan},
title = {{Recent Increments in Incremental View Maintenance}},
year = {2024},
isbn = {9798400704833},
publisher = {Association for Computing Machinery},
address = {New York, NY, USA},
doi = {10.1145/3635138.3654763},
booktitle = {Companion of the 43rd Symposium on Principles of Database Systems},
pages = {8–17},
numpages = {10},
location = {Santiago AA, Chile},
series = {PODS '24}
}

@article{mami2012,
  title={A survey of view selection methods},
  author={Mami, Imene and Bellahsene, Zohra},
  journal={{ACM Sigmod Record}},
  volume={41},
  number={1},
  pages={20-29},
  year={2012},
  publisher={ACM New York, NY, USA}
}

@inproceedings{wang2019,
  title     = {Hyperscan: A Fast Multi-pattern Regex Matcher for Modern CPUs},
  author    = {Xiang Wang and Yang Hong and Harry Chang and KyoungSoo Park and Geoff Langdale and Jiayu Hu and Heqing Zhu},
  booktitle = {Proceedings of the 16th USENIX Symposium on Networked Systems Design and Implementation (NSDI '19)},
  year      = {2019},
  pages     = {631-648}, 
  publisher = {USENIX Association},
  address   = {Boston, MA, USA},
  month     = feb,
  url       = {https://www.usenix.org/system/files/nsdi19-wang-xiang.pdf},
  isbn      = {978-1-931971-49-2}
}

@inproceedings{Korber21,
  author    = {Michael Körber and Nikolaus Glombiewski and Bernhard Seeger},
  title     = {Index-Accelerated Pattern Matching in Event Stores},
  booktitle = {Proceedings of the 2021 ACM SIGMOD International Conference on Management of Data},
  pages     = {1023-1036},
  year      = {2021},
  publisher = {ACM},
  address   = {Virtual Event, China},
  doi       = {10.1145/3448016.3457245}
}

\end{document}